\documentclass[fleqn,usenatbib]{mnras}

\usepackage{newtxtext,newtxmath}

\usepackage{amsmath}	

\usepackage[T1]{fontenc}
\usepackage{ae,aecompl}

\usepackage{graphicx}	

\usepackage{amssymb}	
\usepackage{booktabs}
\usepackage{array}
\usepackage{ulem}
\usepackage{xcolor}

\title[Wind properties of MW and SMC massive stars]
{Wind properties of Milky Way and SMC massive stars:  empirical $Z$ dependence from CMFGEN models}

\author[Marcolino et al.]{
W. L. F. Marcolino$^{1}$\thanks{E-mail: wagner@astro.ufrj.br},
J. -C. Bouret$^{2}$, 
H. J. Rocha-Pinto$^{1}$, 
M. Bernini-Peron$^{1,3}$,  
\newauthor
and J. S. Vink$^{4}$
\\
$^{1}$Observat\'orio do Valongo, Universidade Federal do Rio de Janeiro, Ladeira Pedro Ant\^onio, 43, CEP 20080-090, Rio de Janeiro, Brazil\\
$^{2}$Aix-Marseille Univ, CNRS, CNES, LAM, Marseille, France\\
$^{3}$Zentrum f{\"u}r Astronomie der Universit{\"a}t Heidelberg, Astronomisches Rechen-Institut, M{\"o}nchhofstr. 12-14, 69120 Heidelberg, Germany \\
$^{4}$ Armagh Observatory and Planetarium, College Hill, Armagh, BT61 9DG, Northern Ireland, UK}

\date{Accepted XXX. Received YYY; in original form ZZZ}

\pubyear{2021}

\begin{document}
\label{firstpage}
\pagerange{\pageref{firstpage}--\pageref{lastpage}}
\maketitle
\begin{abstract}
{Detailed knowledge about stellar winds and evolution at different metallicities  
is crucial for understanding stellar populations and feedback in the Local Group of galaxies 
and beyond. Despite efforts in the literature, we still lack a comprehensive, empirical view 
of the dependence of wind properties on metallicity ($Z$). Here, we investigate 
the winds of O and B stars in the Milky Way (MW) and Small Magellanic 
Cloud (SMC). We gathered a sample of 96 stars analyzed by means of the NLTE code CMFGEN. 
We explored their wind strengths and terminal velocities to address the $Z$ dependence, 
over a large luminosity range. The empirical wind-luminosity relation (WLR) obtained updates and extends 
previous results in the literature. It reveals a luminosity and $Z$ dependence, in agreement 
with the radiatively driven wind theory. For bright objects ($\log L/L_\odot \gtrsim 5.4$), 
we infer that $\dot{M} \sim Z^{0.5-0.8}$. However, this dependence seems to get weaker or 
vanish at lower luminosities}. The analysis of the terminal 
velocities suggests a shallow $Z^n$ dependence, with $n \sim 0.1-0.2$, but it should be confirmed with a 
larger sample and more accurate $V_{\infty}$ determinations. Recent results 
on SMC stars based on the PoWR code support our inferred WLR. On the other hand,
recent bow-shocks measurements stand mostly above our derived WLR. 
Theoretical calculations of the WLR are not precise, specially at low $L$, where the results   
scatter. Deviations between our results and recent predictions are identified to be 
due to the weak wind problem and the extreme terminal velocities predicted by the models.
The Z dependence suggested by our analysis deserves further investigations, given its astrophysical 
implications.
\end{abstract}

\begin{keywords}
stars: fundamental parameters -- stars: massive -- stars: mass-loss -- stars: atmospheres -- stars: winds, outflows
\end{keywords}


\section{Introduction}
\label{sec:intro}

The winds of massive stars are mostly driven by the transference of momentum 
from their intense UV radiation field to metal lines \citep*[][]{CAK75, Lamers99}. 
A late-type O star at solar metallicity for example, has its wind driven mainly by Fe ($\sim 23$\%), 
Si ($\sim 20$\%), C ($\sim 14$\%), Al ($\sim 13$\%) and other metal (few \%) lines. H and He 
correspond only to $\sim 13$\% and $\sim 0.05$\% of the radiative force, respectively  
\citep{Mokiem07}. Due to this fact, massive stars in lower metallicity environments are expected - and 
in fact observed - to have weaker winds \citep*[][]{Vink01, Mokiem07}.
These winds represent a significant amount of mass lost even during the relatively short 
lifetime of massive stars \citep*[][]{Ekstrom12}. Moreover, they have an enormous impact 
in the ISM through mechanical energy and chemical elements deposition \citep*[][]{Abbott82}.

In addition to their metallicity ($Z$) dependence, stellar winds are coupled with rotation. 
The mass-loss rate can be affected by rotation and at the same time, remove angular 
momentum, changing the internal rotation profile \citep*[][]{Meynet00}. 
In contrast with low-mass stars ($M \lesssim 8$M$_{\odot}$), single massive star evolution 
is thus a complex function of the initial mass, $Z$, mass-loss, and rotation. 
Such dependence is challenging to describe and implement in models, 
but several efforts have been made in the literature \citep*[e.g.][]{Brott11,Ekstrom12,Georgy13}. 
In fact, if we aim to understand topics such as stellar populations and feedback at low and 
high-redshifts, production of local pair instability supernovae \citep*[e.g.][]{Whalen14}, or even the 
origin of the masses of the binary components that produce gravitational waves detected 
by LIGO \citep*[see e.g.][]{Abbott16}, we must have also a quantitative understanding of stellar evolution and 
winds at different metallicities. 

From an observational point of view, \citet{Mokiem07} were the first to present a detailed 
comparison of wind strengths of massive OB stars at different metallicities. In particular, 
they analyzed several data from the Milky Way (MW), Large Magellanic Cloud (LMC), and Small 
Magellanic Cloud (SMC). A clear mass-loss rate dependence on $Z$ was observed and 
the following relation was inferred: $\dot{M} \propto Z^{0.83 \pm 0.16}$, 
in agreement with the theoretical predictions of \citet{Vink01} within the error bars, namely, 
$\dot{M} \propto Z^{0.69 \pm 0.10}$. This agreement provided a solid framework for the use of the 
theoretical mass-loss recipes in stellar evolution models. However, recent works by \citet{Bjork21} and 
\citet{VinkSander21} for example, provide different dependencies, $\dot{M} \propto Z^{0.95}$ and 
$\dot{M} \propto Z^{0.42}$, respectively, revealing mismatch among theoretical calculations.
To date, the work of \citet{Mokiem07} is still the most comprehensive one on the empirical 
metallicity dependence of the winds of massive OB stars.

Newer high-resolution, multi-wavelength data of hundreds of galactic and extragalactic O and B 
stars were analyzed during the last decade. Different codes and methods were applied to infer 
the stellar and wind parameters of these objects, addressing important questions for stellar evolution and wind theory.  
Fast, robust, stellar atmosphere analyses of large observational datasets were performed 
\citep*[e.g.][]{Sabin17, Agudelo17}, but often at the expense of wavelength coverage (e.g., only optical). 
In parallel, detailed, computational demanding analyses of multi-wavelength data of much smaller samples 
were carried out with co-moving frame, full metal line-blanketing calculations 
\citep*[e.g.][]{Bouret13, Marcolino17, Bouret21}. Despite all these efforts in the 
Milky Way and in the Magellanic Clouds, we still lack a complete, 
updated view of the empirical dependence of wind properties on Z. Part of the problem is related to the 
lack of quantitative UV analyses of low luminosity O and B stars, not taken into account in \citet{Mokiem07} 
and other studies, as we will further discuss.

Our aim in the present paper is to provide an updated, improved view of the metallicity dependence of the winds of O and B stars. To do so, we gathered the results of the analysis of 96 O and B stars, either in the MW or in the SMC, all performed with the code CMFGEN \citep{Hillier98}. This allows us to achieve a robust internal consistency, in the sense that the physical properties of the stars were measured with the same code, under the same basic physical ingredients and assumptions (see below).

We explore wind-luminosity relation (WLR) diagrams and discuss the results in terms of 
the MW and SMC metallicity. We analyze the $Z$ dependence in detail over a large luminosity range. 
We also analyze the terminal velocity ($V_\infty$) of the stars of our sample. 
We estimated the $m$ and $n$ exponents in the $V_\infty \propto $\,Z$^n$ and 
$\dot{M} \propto $\,Z$^m$ relations, which are widely used in the literature. We also 
compare our results with other techniques and measurements 
for O and B stars in the MW and SMC. 

In Section \ref{sec:mdotz} we describe our sample and present our WLR analysis for the MW and SMC stars. 
The metallicity dependence of the mass-loss rate is presented later in Section \ref{sec:mathZ}. 
In Section \ref{sec:vinfz}, the terminal velocities in the MW and SMC are presented along with 
a $V_\infty (Z)$ analysis. In Section \ref{sec:discussion}, we discuss our results, possible 
caveats and other measurements in the literature (e.g., data of SMC stars obtained with the PoWR code 
and bow-shocks measurements of MW stars). We also compare our empirical data with theoretical results 
to check for trends and highlight possible problems. Finally, in Section \ref{sec:conclusions} we 
present a summary of our main findings.

{\renewcommand{\arraystretch}{1.005}
\begin{table*}
\centering 
\begin{tabular}{l c c c c c c c c} 
\toprule
\textbf{Star} & SpT & $\log L/L_\odot$ & $T_{eff}$ (kK) & $R/R_\odot$ & $\log \dot{M}$ ($\log \dot{M}_{H\alpha}$) & $V_\infty$ (km s$^{-1}$) & $\log D$ ($\log D_{H\alpha}$) & Reference \\ 
\midrule 
HD~16691	&	O4If 		    &   $5.94 \pm 0.1$	&   41.0	&   18.66	&	    -4.91	&	2300	&	$29.89 \pm 0.06$ & B12 \\
HD~66811	&	O4I   		    &   $5.91 \pm 0.1$	&   40.0	&   18.94	&	    -5.05	&	2300	&	$29.75 \pm 0.06$ & B12 \\
HD~190429A  &	O4If		    &   $5.96 \pm 0.1$  &   39.0	&   21.10	&	    -4.98	&  	2300	&	$29.84 \pm 0.05$ & B12 \\	     
HD~15570	&   O4If		    &   $5.94 \pm 0.1$	&   38.0	&   21.72	&	    -5.01	&	2200	&	$29.80 \pm 0.05$ & B12 \\
HD~14947	&   O4.5If		    &   $5.83 \pm 0.1$	&   37.0	&   20.19	&	    -5.09	&	2300	&	$29.72 \pm 0.07$ & B12 \\
HD~210839	&   O6I  		    &   $5.80 \pm 0.1$	&   36.0	&   20.60	&	    -5.20	&	2100	&	$29.58 \pm 0.09$ & B12 \\
HD~163758	&   O6.5If   	    &   $5.76 \pm 0.1$	&   34.5	&   21.42	&	    -5.15	&	2100	&	$29.64 \pm 0.08$ & B12 \\
HD~192639	&   O7.5Iabf	    &   $5.68 \pm 0.1$	&   33.5	&   20.72	&	    -5.27	&	1900	&	$29.47 \pm 0.10$ & B12 \\

HD~188001   &   O7.5Iaf         &   $5.69 \pm 0.20$    &   33.0    &   21.60   &       -5.23   &   1800    &  $29.49 \pm 0.41$  & M17 \\
HD~207198   &   O8.5II          &   $5.05 \pm 0.26$    &   32.5    &   10.66   &       -7.00   &   2000    &  $27.61 \pm 0.41$  & M17 \\
HD~30614    &   O9.5Iab         &   $5.81 \pm 0.25$    &   29.0    &   32.34   &       -5.12   &   1600    &  $29.64 \pm 0.41$  & M17 \\
HD~188209   &   O9.5Iab         &   $5.65 \pm 0.26$    &   30.0    &   25.30   &       -5.75   &   2000    &  $29.05 \pm 0.41$  & M17 \\
HD~209975   &   O9.5Ib          &   $5.35 \pm 0.30$    &   30.5    &   17.10   &       -6.50   &   2000    &  $28.22 \pm 0.41$  & M17 \\
HD~195592   &   O9.7Ia          &   $5.47 \pm 0.25$    &   28.0    &   23.29   &       -5.14   &   1400    &  $29.49 \pm 0.41$  & M17 \\

HD~91969	&	B0Ia	        &   $5.52 \pm 0.25$    &   27.5  &	25.3	&       (-6.00)   &   1470	&	($28.67 \pm 0.15$) & C06 \\
HD~94909	&	B0Ia	        &   $5.49 \pm 0.25$    &   27.0	&	25.5	&       (-5.70)   &   1050	&	($28.62 \pm 0.15$) & C06 \\
HD~122879	&   B0Ia	        &   $5.52 \pm 0.25$    &   28.0	&	24.4	&       (-5.52)   &   1620	&	($29.18 \pm 0.15$) & C06 \\
HD~38771	&	B0.5Ia	        &   $5.35 \pm 0.25$    &   26.5	&	22.2	&       (-6.05)   &   1525	&	($28.61 \pm 0.15$) & C06 \\
HD~115842	&   B0.5Ia	        &   $5.65 \pm 0.25$    &   25.5	&	34.2	&       (-5.70)   &   1180	&	($28.94 \pm 0.15$) & C06 \\
HD~152234	&   B0.5Ia	        &   $5.87 \pm 0.25$    &   26.0	&	42.4	&       (-5.57)   &   1450	&	($29.21 \pm 0.15$) & C06 \\

HD~192660   &   B0Ib            &   $5.74 \pm 0.13$   &   30.0  &   23.4    &       (-5.30)   &   1850  &   ($29.45^{+0.00}_{-0.40}$) & S08  \\
HD~204172   &   B0.2Ia          &   $5.48 \pm 0.27$   &   28.5  &   22.4    &       (-6.24)   &   1685  &   ($28.46^{+0.34}_{-0.40}$) & S08 \\
HD~185859	&   B0.5Ia	        &   $5.54 \pm 0.14$   &   26.0 	&   29.1	&       (-6.30)   &   1830	&	($28.49^{+0.08}_{-0.10}$) & S08 \\
HD~213087	&   B0.5Ib	        &   $5.69 \pm 0.11$   &   27.0  &   32.0	&       (-6.15)   &   1520	& 	($28.58^{+0.20}_{-0.00}$) & S08 \\
HD~64760	&	B0.5Ib	        &   $5.48 \pm 0.26$   &   28.0	&   23.3    &       (-5.96)   &   1600	&  	($28.73^{+0.27}_{-1.04}$) & S08 \\

$\epsilon$ Ori & B0Iab          &   $5.60 \pm 0.33$    &   27.5    &   28.0    &       -5.60   &   1800    &  $29.18 \pm 0.22$  & M15a   \\
HD~167264   &   B0.5Iab         &   $5.65 \pm 0.27$    &   28.0    &   28.6    &       -6.00   &   2000    &  $28.83 \pm 0.21$  & M15a    \\

HD~156292	&   O9.7III		    &   $5.12 \pm 0.20$ & 31.0 & 13.0 &       -8.32   &   1300    &   $26.15^{+0.56}_{-0.48}$ & A19 \\
HD~24431    &   O9III           &   $5.17 \pm 0.20$ & 33.0 & 11.9 &  -8.10 (-6.27)  &   2300    &   26.60 ($28.12^{+0.49}_{0.49}$) & A19 \\
HD~105627   &   O9III           &   $5.17 \pm 0.20$ & 33.0 & 11.9 &       -7.89   &   2100    &   $26.74^{+0.72}_{-0.40}$ & A19 \\  
HD~116852   &   O8.5II-III((f)) &   $5.33 \pm 0.20$ & 32.5 & 14.7 &       -6.72   &   2100    &   $27.98^{+0.62}_{-0.60}$ & A19 \\
HD~153426   &   O8.5III         &   $5.24 \pm 0.20$ & 32.0 & 13.7 &  -7.85 (-6.35)  &   2400    &   26.90 ($28.39^{+0.52}_{-0.35}$) & A19 \\
HD~218195   &   O8.5IIINstr     &   $5.24 \pm 0.20$ & 33.0 & 12.9 &  -7.49 (-6.39)  &   2000    &   27.16 ($28.27^{+0.52}_{-0.61}$) & A19 \\ 
HD~36861    &   O8III           &   $5.30 \pm 0.20$ & 33.5 & 13.4 &  -7.10 (-6.39)  &   2000    &   27.56 ($28.28^{+0.38}_{-0.92}$) & A19 \\
HD~115455   &   O8III((f))      &   $5.30 \pm 0.20$ & 34.0 & 13.0 &  -7.80 (-6.15)  &   2300    &   26.92 ($28.36^{0.49}_{0.49}$) & A19 \\
HD~135591   &   O8IV((f))       &   $5.10 \pm 0.20$ & 35.0 &  9.7 &       -7.20   &   2100    &   $27.41^{+0.60}_{-1.12}$ & A19 \\ 

HD~193514   &   O7-7.5III       &   $5.65 \pm 0.09$ & 34.5 & 18.7 &       -5.60   &   2190    &  $29.18 \pm 0.48$ &  M15b \\
HD~193682   &   O5III(f)        &   $5.50 \pm 0.09$ & 39.4 & 12.1 &       -5.70   &   2650    &  $29.06 \pm 0.48$ &  M15b \\
HD~190864   &   O6.5III(f)      &   $5.35 \pm 0.14$ & 38.0 & 10.9 &       -6.40   &   2250    &  $28.27 \pm 0.48$ &  M15b\\  
HD~191978   &   O8III           &   $5.35 \pm 0.23$ & 33.2 & 14.3 &       -8.70   &   1600    &  $25.88 \pm 0.48$ &  M15b \\     

HD~216898   &   O9IV-O8.5V      &   $4.72 \pm 0.25$ & 34.0 & 6.7  &       -9.35  &    1700    &  $25.09 \pm 0.71$ &  M09    \\
HD~326329   &    O9V            &   $4.74 \pm 0.10$ & 31.0 & 8.0  &       -9.22  &    1700    &  $25.26 \pm 0.71$ &  M09  \\
HD~66788    &   O8-9V           &   $4.96 \pm 0.25$ & 34.0 & 8.7  &       -8.92  &    2200    &  $25.69 \pm 0.71$  &  M09  \\
$\zeta$ Oph &   O9.5Vnn         &   $4.86 \pm 0.10$ & 32.0 & 9.2  &       -8.80  &    1500    &  $25.66 \pm 0.71$  &  M09 \\
HD~216532   &   O8.5V((n))      &   $4.79 \pm 0.25$ & 33.0 & 7.5  &       -9.22  &    1500    &  $25.19 \pm 0.72$ &  M09 \\

HD~46223    & O4V((f))          &  $5.60 \pm 0.11$  &   43.0    &  11.47    &   -6.67 (-5.70) & 2800 & 28.11 ($29.08 \pm 0.48$) & M12 \\
HD~46150    & O5V((f))z         &  $5.65 \pm 0.25$  &   42.0    &  12.73    &   -6.80 (-5.90) & 2800 & 28.00 ($28.90 \pm 0.48$) & M12 \\
HD~46485    & O7Vn              &  $5.05 \pm 0.11$  &   36.0    &  8.69     &   -7.80 (-6.45) & 1850 & 26.74 ($28.09 \pm 0.48$) & M12 \\
HD~46202    & O9.5V             &  $4.85 \pm 0.12$  &   33.5    & 7.97      &   -9.00 (-7.10) & 1200 & 25.33 ($27.23 \pm 0.48$) & M12 \\
HD~48279    & ON8.5V            &  $4.95 \pm 0.11$  &   34.5    & 8.43      &   -8.80 (-6.80) & 1300 & 25.58 ($27.58 \pm 0.48$) & M12 \\
HD~46966    & O8.5IV            &  $5.20 \pm 0.11$  &   35.0    & 10.92     &   -8.00 (-6.40) & 2300 & 26.68 ($28.28 \pm 0.48$) & M12 \\

HD~38666    & O9.5V             & $4.66^{+0.40}_{-0.30}$  &   33.0    & 6.58      &       -9.50   & 1200      & $24.79 \pm 0.71$ & M05 \\
HD~34078    & O9.5V             & $4.77^{+0.41}_{-0.32}$  &   33.0    & 7.47      &       -9.50   & 800       & $24.64 \pm 0.72$ & M05 \\
HD~93028    & O9V               & $5.05 \pm 0.22$      &   34.0    & 9.71      &       -9.00   & 1300      & $25.41 \pm 0.71$ & M05 \\
HD~152590   & O7.5Vz            & $4.79^{+0.33}_{-0.24}$      &   36.0    & 6.42      &       -7.78   & 1750      & $26.67 \pm 0.71$ & M05 \\
HD~93146    & O6.5V((f))        & $5.22^{+0.23}_{-0.25}$      &   37.0    & 9.97      &       -7.25   & 2800      & $27.50 \pm 0.70$ & M05 \\
HD~42088    & O6.5Vz            & $5.23 \pm 0.19$      &   38.0    & 9.56      &       -8.00   & 1900      & $26.57 \pm 0.70$ & M05 \\
HD~93204    & O5V((f))          & $5.51^{+0.25}_{-0.20}$      &   40.0    & 11.91     &       -6.25   & 2900      & $28.55 \pm 0.70$ & M05 \\
HD~15629    & O5V((f))          & $5.56 \pm 0.18$      &   41.0    & 12.01     &       -6.00   & 2800      & $28.79 \pm 0.70$ & M05 \\
HD~93250$^\dagger$    & O3.5V((f+))       & $6.12^{+0.25}_{-0.17}$      &   44.0    & 19.87     &   -5.25   & 3000 & $29.68 \pm 0.70$ & M05 \\
\bottomrule 
\end{tabular}
\caption{Stellar and wind parameters of Milky Way O and B stars analyzed with CMFGEN models. Homogeneous wind parameters are shown. 
Clumped models had the mass-loss rates re-scaled with a $\dot{M}/\sqrt{f}$ factor (see text). 
References: B12 = \citet{Bouret12}. M17 = \citet{Marcolino17}. C06 = \citet{Crowther06}. S08 = \citet{Searle08}. M15a = 
\citet{Martins15}. A19 = \citet{Elisson19}. M15b = \citet{Mahy15}. M09 = \citet{Marcolino09}. 
M12 = \citet{Martins12}. M05 = \citet{Martins05}. $^\dagger$This star is mentioned as a prototype of the O3.5V class 
in \citet{Walborn02}. However, the Galactic O Star Catalog of \citet{Maiz16} reports an O4IV(fc) classification.} 
\label{tab:sampleMW} 
\end{table*}}


\begin{table*}
\centering 
\begin{tabular}{l c c c c c c c c} 
\toprule
\textbf{Star} & SpT & $\log L/L_\odot$ & $T_{eff}$ (kK) & $R/R_\odot$ & $\log \dot{M}$ ($\log \dot{M}_{H\alpha}$) & $V_\infty$  (km s$^{-1}$) & $\log D$ ($\log D_{H\alpha}$) & Reference \\ 
\midrule 
AzV~75       &   O5.5I(f)        &  $5.94 \pm 0.10$     &   38.5   &   21.16    &       -5.80   &   2050    &  $28.97 \pm 0.20$ & B21 \\
AzV~15       &     O6.5I(f)      &  $5.83 \pm 0.10$     &   39.0   &   18.17     &      -5.96   &   2050    &  $28.78 \pm 0.20$ & B21 \\
AzV~232      &     O7Iaf+        &  $5.89 \pm 0.10$     &   33.5   &   26.39     &      -5.34   &   1350    &  $29.30 \pm 0.20$ & B21 \\
AzV~83       &     O7Iaf+        &  $5.54 \pm 0.10$     &   32.8   &   18.40       &    -5.64   &    940    &  $28.77 \pm 0.21$ & B21 \\
AzV~327      &     O9.5II-Ibw    &  $5.54 \pm 0.10$     &   30.0   &       21.99  &    -6.87   &   1500    &   $27.78 \pm 0.20$ & B21 \\ 
MPG~355      &    ON2III         &  $6.04 \pm 0.10$     &   51.7   &        13.17  &     -5.89  &   2800    &  $28.92 \pm 0.05$ & B13  \\
AzV~77       &     O7III         &  $5.40 \pm 0.10$    &   37.5   &        11.98   &   -7.38   &   1400    &  $27.10 \pm 0.20$ & B21 \\
AzV~95       &     O7III((f))    &  $5.46 \pm 0.10$    &   38.0   &        12.50    &  -6.90   &   1700    &  $27.68 \pm 0.20$ & B21 \\
AzV~69       &     OC7.5III((f)) &  $5.61 \pm 0.10$    &   33.9   &        18.67   &     -6.01 &   1800    &  $28.68 \pm 0.20$ & B21 \\
AzV~47       &     O8III((f))    &  $5.44 \pm 0.10$    &   35.0   &       14.40    &    -7.68  &   2000    &  $27.00 \pm 0.20$ & B21 \\
AzV~307      &     O9III         &  $5.15 \pm 0.10$    &   30.0   &        14.04   &    -8.32  &   1300    &  $26.17 \pm 0.20$ & B21 \\
Azv~439      &     O9.5III       &  $5.16 \pm 0.10$    &   31.0   &        13.30   &     -7.35  &   1000    & $27.01 \pm 0.21$ & B21 \\
AzV~170      &     O9.7III       &  $5.14 \pm 0.10$    &   30.5   &        13.43  &     -8.32  &   1200    & $26.12 \pm 0.21$ & B21 \\
AzV~43       &     B0.5III       &  $5.13 \pm 0.10$    &   28.5   &        15.20  &     -7.65  &   1200    & $26.82 \pm 0.21$ & B21 \\
AzV~177      &   O4V((f))        &  $5.43 \pm 0.10$   &   44.5    &   8.81    &       -6.20   &   2400    &  $28.45 \pm 0.05$ & B13 \\
AzV~388      &   O4V             &  $5.54 \pm 0.10$    &   43.1    &   10.65   &       -6.52   &   2100    &  $28.12 \pm 0.05$ & B13 \\
MPG~324      &   O4V             &  $5.51 \pm 0.10$    &   42.1    &   10.79   &       -6.27   &   2300    &   $28.41 \pm 0.05$ & B13 \\
MPG~368      &   O6V             &  $5.38 \pm 0.10$   &   39.3    &   10.66   &       -6.93   &   2100    &   $27.71 \pm 0.05$ & B13 \\
AzV~243      &   O6V             &  $5.59 \pm 0.10$   &   39.6    &   13.37   &       -6.45   &   2000    &   $28.21 \pm 0.05$ & B13 \\
AzV~446	     &	 O6.5V           &	$5.25 \pm 0.10$	 &   39.7    &   8.99	 &	     -7.90   &   1400	 &   $26.52 \pm 0.06$ & B13 \\
AvZ~429		 &   O7V    		 &  $5.13 \pm 0.10$	 &	 38.3    &   8.42	 &   	 -7.90   & 	 1300	 &	 $26.48 \pm 0.06$ & B13 \\
MPG~113	     &   OC6Vz 		     &  $5.15 \pm 0.10$  &   39.6    &   8.06    &       -8.52   &   1250    &   $<25.83$ & B13 \\
MPG~356      &   O6.5V           &  $4.88 \pm 0.10$  &   38.2    &   6.34    &       -8.46   &   1400    &   $<25.89$ & B13 \\
MPG~523      &   O7Vz            &  $4.80 \pm 0.10$   &   38.7    &   5.64    &       -9.22   &   1950    &  $<25.25$ & B13 \\
NGC346-046   &   O7Vn            &  $4.81 \pm 0.10$   &   39.0    &   5.62    &       -9.22   &   1950    &   $<25.24$ & B13 \\
NGC346-031   & 	 O8Vz            &  $4.95 \pm 0.10$   &   37.2    &   7.25    &       -9.22   &   1540    &   $<25.20$ & B13 \\
AzV~267	     &   O8V             &	$4.90 \pm 0.10$	 &   35.7    &	 7.43	 &	     -8.10	 &   1250    &	 $26.23 \pm 0.06$ & B13 \\
AzV~461      &   O8V             &  $5.00 \pm 0.10$   &   37.1    &   7.72    &       -9.00   &   1540    &   $<25.43$ & B13 \\
MPG~299      &   O8Vn            &  $4.64 \pm 0.10$   &   36.3    &   5.33    &       -8.52   &   1540    &   $<25.83$ & B13 \\
MPG~487      &   O8V             &  $5.12 \pm 0.10$   &   35.8    &   9.52    &       -8.52   &   1540    &   $<25.96$ & B13 \\
AzV~468      &	 O8.5V           &  $4.76 \pm 0.10$    &   34.7    &   6.70	 &	     -9.15	 &   1540    &   $<25.25$ & B13 \\
AzV~148      &   O8.5V           &  $4.84 \pm 0.10$    &   32.3    &   8.47    &       -8.70   &   1540    &   $25.75 \pm 0.06$ & B13 \\
MPG~682      &   O9V             &  $4.89 \pm 0.10$   &   34.8    &   7.73    &       -9.05	 &   1250    &   $<25.29$ & B13 \\
AzV~326      &   O9V             &  $4.81 \pm 0.10$   &   32.4    &   8.14    &		 -9.15   &   1250	 &	 $<25.20$ & B13 \\
AzV~189      &   O9V             &  $4.81 \pm 0.10$    &   32.3    &   8.19    &       -9.22   &   1250    &  $<25.13$ & B13 \\
MPG~012	     & 	 B0IV            &  $4.93 \pm 0.10$   &   31.0    &  10.20    &       -9.30   &   1250    & 	$<25.10$ & B13 \\
\bottomrule 
\end{tabular}
\caption{Stellar and wind parameters of Small Magellanic Cloud O and B stars analyzed with CMFGEN models. Homogeneous wind parameters are shown ($f = 1.0$). Clumped models had the mass-loss rates re-scaled with a $\dot{M}/\sqrt{f}$ factor. 
References: B13 = \citet{Bouret13}; B21 = \citet{Bouret21}.} 
\label{tab:sampleSMC} 
\end{table*}


\section{Stellar Wind Strengths - Milky Way and SMC}
\label{sec:mdotz}

\subsection{Observational data}
\label{sec:homogeneity}

In this paper, we focus on a comparison of the wind properties of O and B stars in the 
 MW and SMC, these populations offer the wider span in metallicity.

We selected several studies of O and B stars in both galaxies from the literature 
\citep[][]{Bouret12,Bouret13,Bouret21,Crowther06, Elisson19, Mahy15, Marcolino09, Marcolino17, 
Martins12, Martins15, Searle08}, all relying on detailed joint analysis of UV and optical 
spectra performed with CMFGEN \citep{Hillier98}. Most of them were done by our group during the last decade. 

We present our sample in Tables \ref{tab:sampleMW} and \ref{tab:sampleSMC}. 
In total, we gathered  60 Milky Way stars and 36 SMC stars. Early to late type O stars 
of all main luminosity classes are comprised (I, III, and V). Most B stars are supergiants, 
with the exception of two sub-giants and one giant in the SMC. Details regarding the 
physical parameters present in these tables will be discussed in the next section.

\begin{figure}
	\includegraphics[width=\columnwidth]{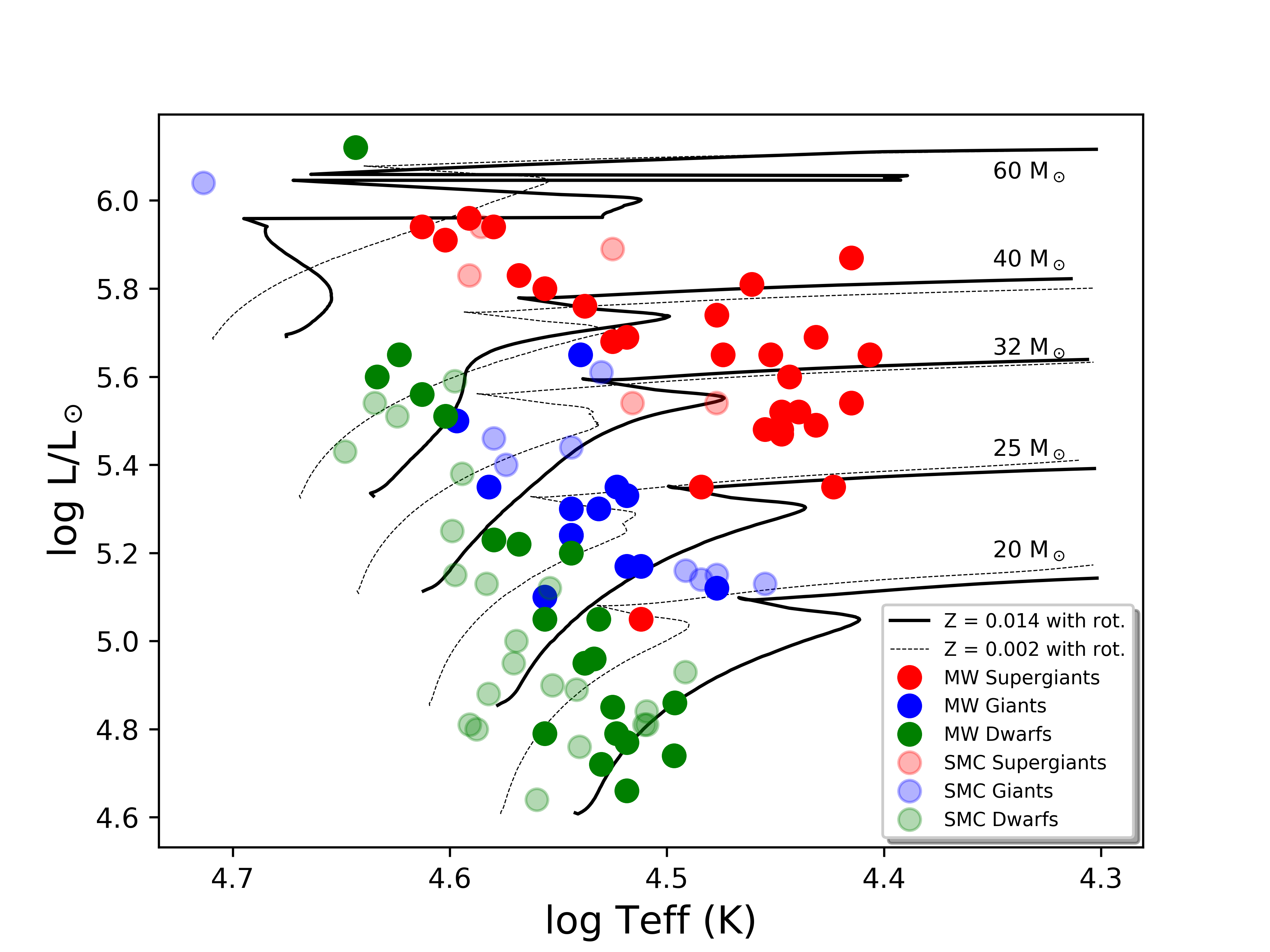}
    \caption{Hertzsprung-Russell diagram for the Milky Way and SMC stars of our sample. Evolutionary tracks from \citet{Ekstrom12} and \citet{Georgy13} are indicated (see text for more details). Note that I-III-V stars 
    occupy different loci in the diagram.} 
    \label{fig:HRdiagram}
\end{figure}

For visualization purposes, we present in Fig. \ref{fig:HRdiagram} the location of these stars 
in the Hertzsprung-Russell (H-R) diagram. We also include evolutionary tracks with 
rotation for Z = 0.014 (solar metallicity) from \citet{Ekstrom12} and Z = 0.002 
(appropriate for SMC) from \citet{Georgy13}. These tracks follow from models with the same physical 
ingredients, e.g., $V_{ini}/V_{crit} = 0.4$, convection parameters, and nuclear reaction rates 
\citep[for more details, see][]{Georgy13}. A large interval in effective temperature and luminosity 
is covered and there is a fair number of stars in each of the luminosity classes (I, III, and V). 
Moreover, as expected, the different classes occupy different loci in the H-R diagram.

By selecting studies performed with the same atmospheric code, which are based on the same basic 
set of physical assumptions, we ensure that the results we analyse have internal consistency. 
On the other hand, we acknowledge that systematics and biases related to the details of the 
spectroscopic analysis methodology and what data were available for the analysis, are impossible 
to grasp at this point.

For example, the CMFGEN models used present a certain degree of diversity. Some works do not
adopt the same atomic data (i.e they use different ion models with various number of transitions, 
energy levels, and superlevels), or they do not consider the same set of chemical elements, or 
use a different version of the code (this last point should not be an issue). In addition, although 
the effective set of sensitive spectral lines useful to characterise the photospheric and wind 
properties of O and B stars is common knowledge \citep[see e.g.,][]{martins11}, the selected 
studies used less or more diagnostics than others for a specific parameter, 
depending on the availability and/or quality of the data. Ideally, a tailored re-analysis of the 
whole sample should be done with fully homogeneous modelling assumptions and new observational data, but this beyond 
the scope of the paper.

For galactic stars, most works (see refs above) relied on UV data from the IUE and 
FUSE satellites. IUE observations are mainly from the short wavelength spectrograph (SWP) 
($\sim 1150-2000$\AA), with resolving power R $\sim 10000$. FUSE spectra are mainly from 
the LiF2A channel ($\sim 1086-1183$\AA), with R $\sim$ 20000. This channel contains the 
\mbox{PV} $\lambda\lambda$1118, 1128  doublet, whose behaviour lead \citet{Fullerton06} 
to uncover the P v problem, i.e the discordance between mass-loss rates derived from P v 
with those obtained from H$\alpha$, which subsequently triggered an extensive research 
on wind clumping as a solution of this problem \citep[see][for a review]{Puls08}. Note 
that Copernicus data were also used for a few Galactic stars \citep[e.g.,][]{Marcolino09, Bouret12}, 
when FUSE data were not available.

For SMC stars, the HST observations were acquired with COS (Cosmic Origins Spectrograph, 
R $\sim 20000$) or STIS (Space Telescope Imaging Spectrograph, R $\sim 40000$). FUSE data 
are also available for some stars.

Most optical observations are also of high-resolution (with resolving power R $\gtrsim$ 40000) 
but intermediate-resolution (resolving power R $\sim 6000-10000$) was also 
used \citep*[mostly for B supergiants, see][]{Crowther06, Searle08}. There are only 
a few exceptions where either the optical spectrum is missing or it is 
obtained at relatively low spectral resolving power (R $\sim$ 2000; 2dF spectra) \citep*[e.g. in][]{Bouret13, Bouret21}. 
Note that for some stars the mid-IR were also modelled \citep{Marcolino17}. This multi-wavelength 
aspect is fundamental to measure values of physical parameters of interest with the highest fidelity. 

UV spectra are notoriously important when studying the wind properties of hot massive stars 
because they contain several resonance doublets that provide the best diagnostics to derive 
the mass-loss rates and wind terminal velocities (not accessible in the optical). The 
sensitivity of the most important diagnostic wind line accessible to optical spectroscopy, 
namely H$\alpha$, is limited to winds with mass loss rates greater than about 
10$^{-7}$ M$_\odot$ yr$^{-1}$ \citep{Marcolino09}. Winds 100 to 1000 times weaker can 
however be measured from the several UV resonance doublets and excited state wind lines, 
which is especially important for low-metallicity O stars, including those with weak winds 
\citep{Martins05, Marcolino09}. 

In general, it is possible to constrain the intrinsic stellar luminosity by comparing 
the theoretical spectral energy distribution (SED) predicted by a model for a set of fundamental
parameters (mostly $T_{eff}$, log $g$, R, $\dot{M}$, Z) to the observed multi-wavelength data 
when extinction amounts and laws, as well as a distance, are taken into account. Distances 
to SMC stars are better constrained, in the sense that it can be considered they all share 
the same distance modulus of the SMC \citep*[e.g., DM = 18.91$\pm 0.02$;][]{harries03}. 
They often suffer minimal extinction, leading to reliable estimates of stellar luminosities, 
radii and masses, crucial parameters in modelling their winds. This is not the case for Galactic
stars, where distances are both uncertain and varied, and extinction can be significant. 
In this case, either distances are adopted and the luminosities then derived from a SED 
fit, or the luminosities are fixed and spectroscopic distances can be inferred, which are 
then checked against observed parallaxes, if available. Although both strategies are 
reasonable and have been used for some stars in our sample \citep*[see e.g.,][]{Martins12, Elisson19}, 
uncertainties on quantities later used in this paper, namely stellar radii and luminosities, are 
notoriously higher for Galactic stars than for their SMC counterparts. 

\subsection{Consistent mass-loss rates: UV and optical}
\label{sec:consistent}

 We will start with the wind momentum-luminosity relation, the WLR diagram: 
 $\log D_{mom}$ -- $\log L/L_\odot$, where $L$ is the bolometric luminosity. 
 By definition, $D_{mom} = \dot{M}V_\infty \sqrt{R_\star}$.  This quantity is 
 called modified momentum and for theoretical reasons it is expected to correlate 
 well with the luminosity  \citep*[for more details, see][]{Kudritzki00}. 
 In this relation, we used unclumped values for the mass-loss rates throughout the paper. If a model 
 is clumped, the mass-loss rate is recomputed according 
 to $\dot{M}_{uncl} =  \dot{M}_{cl}/\sqrt{f}$ \citep*[f = micro-clumping parameter; see][]{Bouret05}.

 
\begin{figure*}
	\includegraphics[width=\textwidth]{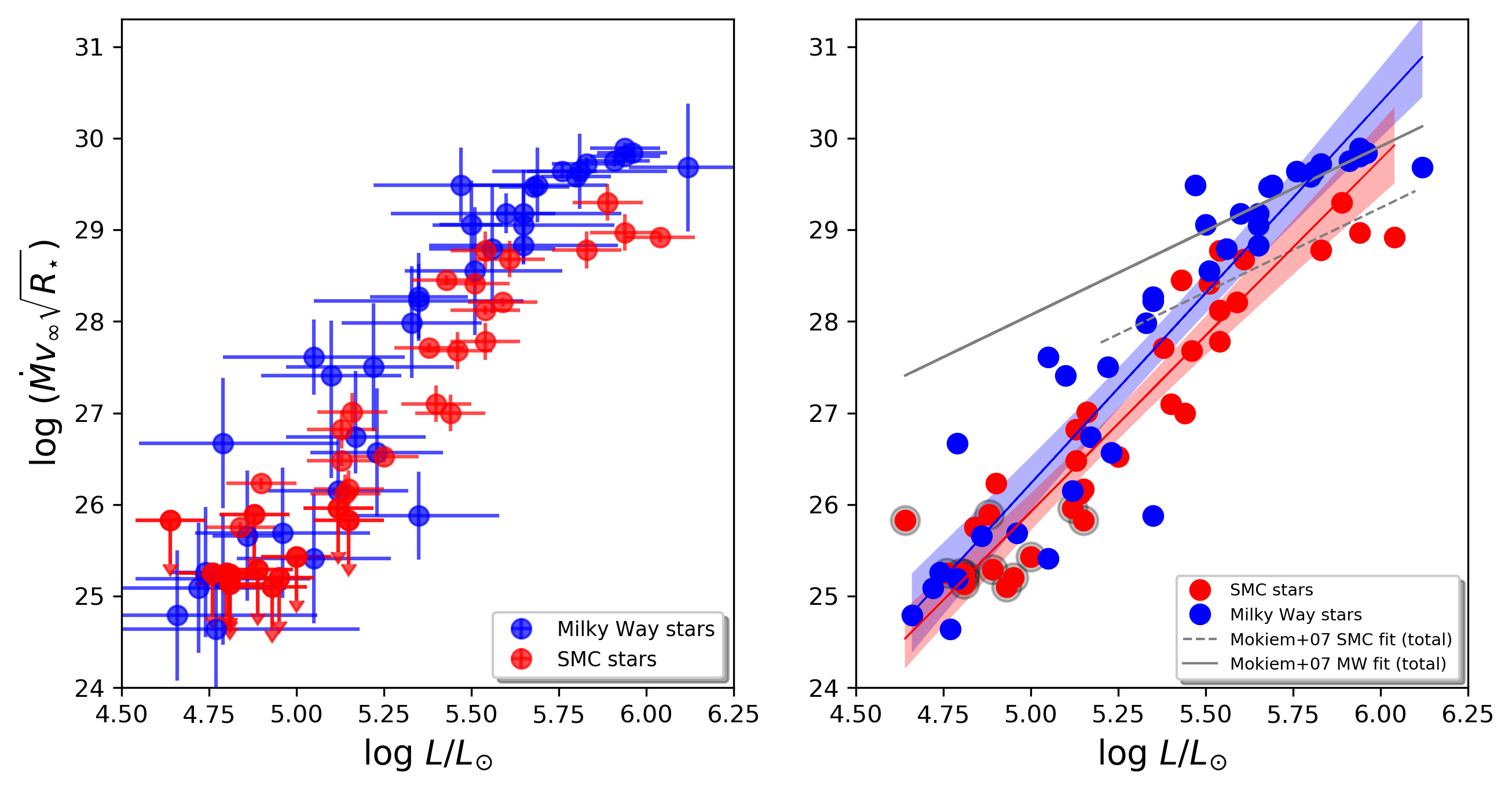}
    \caption{Wind momentum-luminosity relation of Milky Way and Small Magellanic Cloud O and B stars. For clarity, the left panel presents 
only the data and the respective error bars. The right panel shows these same points along with regression lines and 2-$\sigma$ 
confidence bands (see text for more details). Note that for $\log L/L_\odot \lesssim 5.4$ the MW and SMC data start to overlap. Only CMFGEN results 
based on a multiwavelength analysis are considered.}
    \label{fig:MWSMC}
\end{figure*}
 
In this section, we use results from models that successfully fitted ultraviolet and 
optical data at the same time, i.e., that refer to a single mass-loss rate for an object. 
The parameters used are listed in Tables \ref{tab:sampleMW} and \ref{tab:sampleSMC}. 
Mass-loss rate values between parentheses were obtained from the H$\alpha$ line, independent of the UV. 
Stars with $\dot{M}$(H$\alpha$) are not used in this section (but see next section). 
We excluded the star HD~191423 of \citet{Mahy15} from the analysis. This star has an 
atypical $V_{\infty}$ of 600 km s$^{-1}$ and very fast rotation, $V \sin I \sim 400$ km s$^{-1}$. 

We present the WLR in Fig.~\ref{fig:MWSMC}. The left panel shows only the data and 
the right panel presents them along with linear fits and confidence bands (more details below). 
To compute the uncertainty on the modified momentum, we need the uncertainties on 
$\dot{M}$, $V_\infty$, and radius. However, they are not present in some of the 
works\footnote{For example, \citet{Searle08} do not provide errors on $V_\infty$. 
This is also true in \citet{Mahy15} for some objects of their sample, 
where $V_\infty$ values from \citet{Prinja90} were adopted, and in \citet{Crowther06}. 
For some objects, mass-loss rates uncertainties are not reported either, neither radii 
uncertainties \citep[see e.g.,][]{Martins12, Mahy15}. On the other hand, some works 
provide all the necessary information \citep[e.g.,][]{Elisson19}, which we promptly used.} 
quoted in Tables \ref{tab:sampleMW} and \ref{tab:sampleSMC}. Without $V_\infty$ and radii 
errors, we adopted that the uncertainty on $D_{mom}$ was purely due to the mass-loss rate uncertainty. 
Although not ideal, it is a reasonable assumption as both $V_\infty$ and radius are 
usually known within much less than ~50\% -- i.e., the mass-loss rate uncertainty usually dominates 
\citep[see e.g.,][]{Martins05}. However, when not available, the uncertainty on $\dot{M}$ 
was adopted to be a factor of 3. Regarding the luminosity, when not available, 
0.25 dex and 0.10 dex of uncertainty were adopted, which are typical values for Galactic and SMC
stars, respectively.

The fits to our MW and SMC data (Fig. \ref{fig:MWSMC}, right panel) were computed with 
Deming regression, which takes the uncertainties both in $D_{mom}$ and $L$ into 
account \citep[][]{Therneau18}. For comparison, the WLR relations obtained by \citet{Mokiem07} are also 
presented. We also computed confidence bands, which are displayed with 2-$\sigma$ as shaded regions. 
We note that the bands provided in the work of \citet{Mokiem07} are 1-$\sigma$ and therefore, much thinner than 
ours (see their Fig. 4). All coefficients with the respective uncertainties are presented in Table \ref{tab:linearcoeffs}. 
They will be explored later in Section \ref{sec:mathZ}. 

In Fig. \ref{fig:MWSMC}, we observe a clear relation between the modified 
wind momentum and the luminosity, as expected, in agreement with the radiatively driven 
wind theory \citep*[see][]{Kudritzki00}. This is observed for both the Milky Way and SMC set. 
In general, the Milky Way points are above the SMC ones, indicating a metallicity 
dependence. However, there seems to be a metallicity degenerescence at low $L$, 
where MW and SMC points start to overlap.

We applied the Kolmogorov-Smirnov test to check whether the distributions of the Milky Way 
and SMC stars in Fig. \ref{fig:MWSMC} could arise from the same population. We found a $p$-value 
of 0.00657 for the null hypothesis that these two samples come from the same population, i.e., 
overall, we have a metallicity dependence on the wind strength, as the linear fits suggest. 
However, we also applied this test for the stars below $\log L/L_{\odot} = 5.2$. The corresponding 
$p$-value increases to 0.796, supporting a degenerescence.

Despite the facts aforementioned, in Fig. \ref{fig:MWSMC} we tagged a small 
subset of SMC stars (arrows in the left panel and red symbols encircled in grey 
in the right panel). These objects belong to the sample of \citet{Bouret13}. 
For them, the models fit well the observations but P-Cygni features are not 
conspicuous in the UV. The corresponding mass-loss rates were thus considered 
to be upper limits. Without these points, the angular coefficient of 
the fit to the SMC stars naturally decreases. This indicates that an 
analysis of more SMC objects with about the same luminosity values is 
urgently needed. The degeneracy might be weakened, removed or confirmed with 
more data points. Interestingly, two late O type stars - AzV 267 (O8V) and AzV 148 (O8.5V) - 
present conspicuous wind profiles in the UV \citep*[see][]{Bouret13}, allowing 
to derive specific values of their mass-loss rates. Both fall at low $L$ and 
above some MW objects in the WLR, supporting the degeneracy. In fact, it is 
likely that the upper limits indicated in Fig. \ref{fig:MWSMC} are actually 
the true values because we observe a gradual decline of the wind profile intensities 
from early to late-type stars, which is reflected also on a gradual decline of 
the mass-loss rates, instead of jumps of some orders of magnitude (e.g., needed 
for a complete separation of the MW and SMC populations at low $L$).

We also note in Fig. \ref{fig:MWSMC} the discrepancy between our results (UV + optical) and the fits 
by \citet{Mokiem07}. These authors provided the most complete, empirical view of the wind metallicity 
dependence at that time. By analyzing MW, SMC, and LMC data 
of several O and B stars, they found a clear evidence that MW stars have stronger winds 
than LMC stars, which in turn have stronger winds than SMC stars. Although {\it a priori} 
expected by the radiatively driven wind theory, this had never been shown quantitatively 
with such clarity (see their Fig. 4). 

Our data stand mostly below the fits by \citet{Mokiem07}. The slopes    
are very different. The main limitation of their study is the lack of modified 
momentum data for Magellanic Cloud stars with low luminosities. 
They constructed the empirical relations (linear fits) from the most luminous stars only, 
where $\log L/L_\odot \gtrsim 5.2$. Data points at lower $L$ were upper limits that 
they neglected in the analysis. This is the reason for the interrupted dashed line 
in Fig. \ref{fig:MWSMC}, for the SMC. The main cause of this issue is the 
difficulty in obtaining wind parameters with optical data. 
H$\alpha$ is essentially in absorption at low $L$ and often contaminated in stars 
close to nebular regions \citep*[see e.g.,][]{Ramachandran19}. We will come back to this 
question later in Section \ref{sec:discussion}. 

The CMFGEN measurements shown in Fig.~\ref{fig:MWSMC} reveal an improved empirical relation 
with a reasonable number of stars in a large luminosity range  
(from dwarfs to supergiants). It updates and extends the results of \citet{Mokiem07}.
It should be noted that Mokiem's relation is in general followed by the theoretical predictions  
of \citet{Vink01}, with only a small offset ($\sim$ 0.2 dex, without clumping correction). This  
reasonable agreement strongly motivated works in the literature to use the theoretical recipe and 
the inferred Z dependence \citep*[e.g.,][]{Ekstrom12}. However, the results presented 
here show that this can be prone to errors, depending on the luminosity regime.

\subsection{Inclusion of $\dot{M}$(H$\alpha$) results}
\label{sec:halpharesults}

\begin{table}
\centering
\begin{tabular}{l c c} 
 \hline\hline
WLR  & linear coeff. ($\alpha$) & slope ($\beta$) \\ [0.5ex] 
 \hline
 Milky Way (UV+optical)   &   5.43 $\pm$ 1.28  & 4.16 $\pm$ 0.23  \\ 
 SMC       (UV+optical)   &   6.67 $\pm$ 1.52  & 3.85 $\pm$ 0.29  \\ 
 \hline
Milky Way (H$\alpha$)     &   8.71 $\pm$ 1.64  & 3.60 $\pm$ 0.29 \\
SMC (H$\alpha$)           &   6.67 $\pm$ 1.52  & 3.85 $\pm$ 0.29 \\ 
\hline
Milky Way (Mokiem)        &   18.87 $\pm$ 0.98 & 1.84 $\pm$ 0.17 \\
SMC (Mokiem)              &   18.20 $\pm$ 1.09 & 1.84 $\pm$ 0.19 \\[1ex]
\hline
\end{tabular}
\caption{Fit parameters to the MW and SMC data in Figs. \ref{fig:MWSMC} 
(UV+optical; see Section \ref{sec:consistent}) and \ref{fig:Halpha-v1} (emphasis on H$\alpha$; see Section \ref{sec:halpharesults}). 
Fit parameters to the empirical data presented by \citet{Mokiem07} is also shown (see their Table 3).}
\label{tab:linearcoeffs}
\end{table}

\begin{figure}
	\includegraphics[width=\columnwidth]{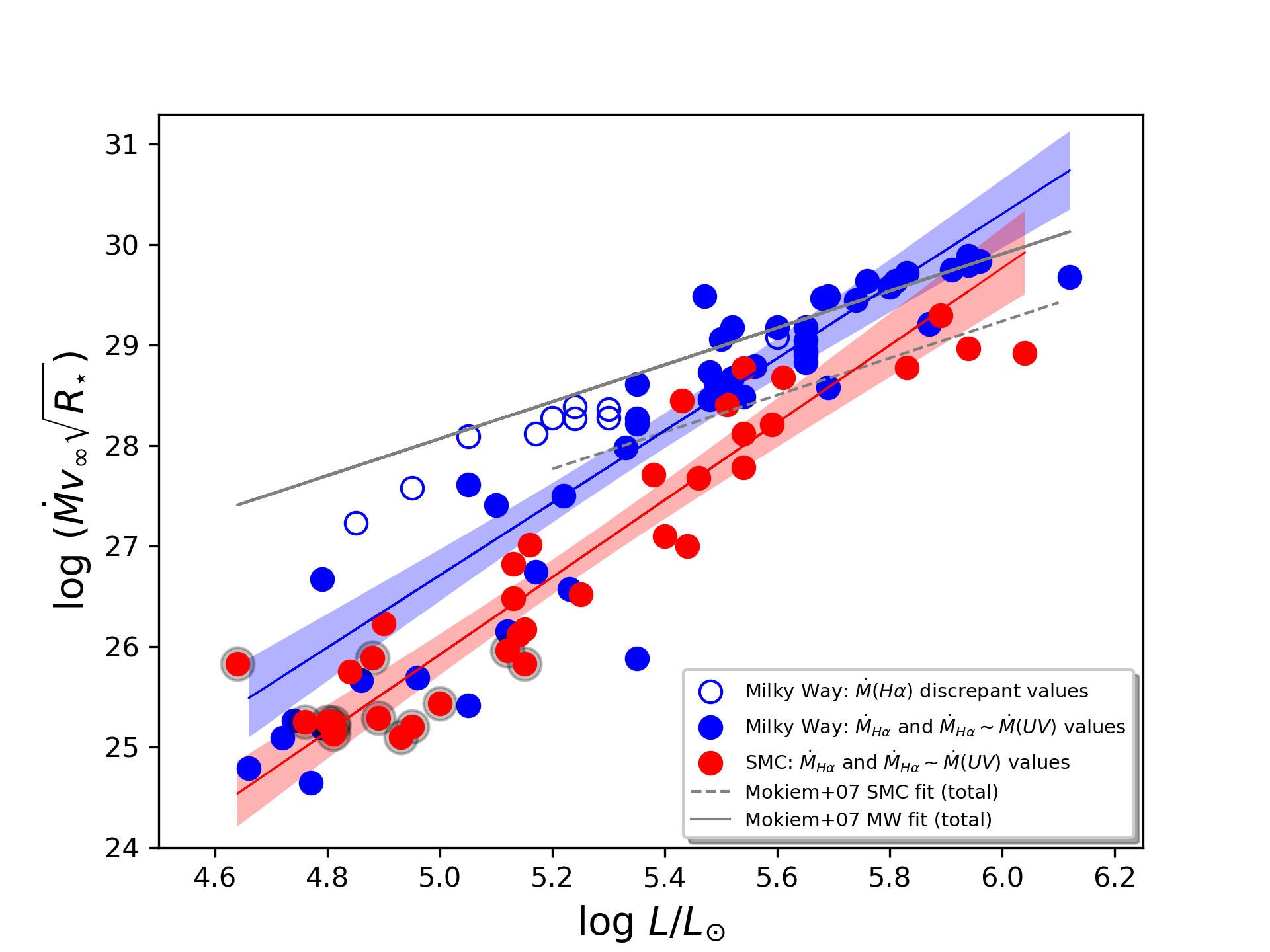}
    \caption{Same as Fig.~\ref{fig:MWSMC} (right panel), but preferring H$\alpha$ over UV mass-loss rates measurements (see text for more details).}
    \label{fig:Halpha-v1}
\end{figure}

In this section, we include mass-loss rates results based on optical -- H$\alpha$, essentially. 
We consider them separately from the previous ones because some represent problematic 
measurements. There is a discrepancy between the $\dot{M}$ value obtained from H$\alpha$ and 
from the UV for a few objects. That is, a single mass-loss rate does not fit the whole spectrum 
of an object, even when clumping and x-rays are taken into account \citep[e.g.,][]{Martins12}. 
This is an open problem and will be discussed later in the paper. Some other results have $\dot{M}$ 
inferred from H$\alpha$ and the fit to the corresponding UV spectra merely checked, with 
important discrepancies in some cases \citep[e.g.,][]{Crowther06}. We include these H$\alpha$ 
results for completeness, as they were also obtained with CMFGEN. In Table \ref{tab:sampleMW}, they are 
denoted between parentheses. In some sense, our analysis anticipates the tendency of the WLR if the 
$\dot{M}$(H$\alpha$) values turn out to be preferred over $\dot{M}$(UV) ones for these objects in 
future works.

We proceed as follows: we keep data from the previous section where a single, robust mass-loss rate 
was obtained, i.e., when $\dot{M}$(H$\alpha$)$\sim \dot{M}$(UV). However, we add: (i) mass-loss 
rates based on fits to H$\alpha$ and (ii) results for $\dot{M}$(H$\alpha$) that diverges from $\dot{M}$(UV). 
The samples used are from \citet{Martins12} (O stars from Monoceros and NGC 2244), \citet{Elisson19} 
(some late O giants), \citet{Crowther06} (early B supergiants only) and \citet{Searle08} (early B supergiants only).

We present the WLR in Fig.~\ref{fig:Halpha-v1}. We first note that the Milky Way stars 
are more scattered in this diagram, specially from mid to low luminosity values, in comparison 
with the previous figure. This reflects the effect of higher mass-loss rates from H$\alpha$ in comparison with the 
UV rates for some objects \citep*[see][]{Martins12}. When one favors H$\alpha$, the tendency is always 
to decrease the WLR slope.


Some of the features observed in Fig.~\ref{fig:MWSMC} are also in Fig.~\ref{fig:Halpha-v1}: 
(i) we have more wind momentum as the luminosity increases, as expected; (ii) Small Magellanic 
Cloud massive stars present weaker winds compared to their Milky Way counterparts; (iii) the linear 
fits still indicate a severe departure from the fits of \citet{Mokiem07} (see also Table \ref{tab:linearcoeffs}).
However, now the degeneracy at low $L$ is not apparent. In order to better address these issues, we explore 
below the metallicity dependence of the mass-loss rate in a quantitative way.

\section{Empirical metallicity dependence}
\label{sec:mathZ}

\begin{figure}
	\includegraphics[width=\columnwidth]{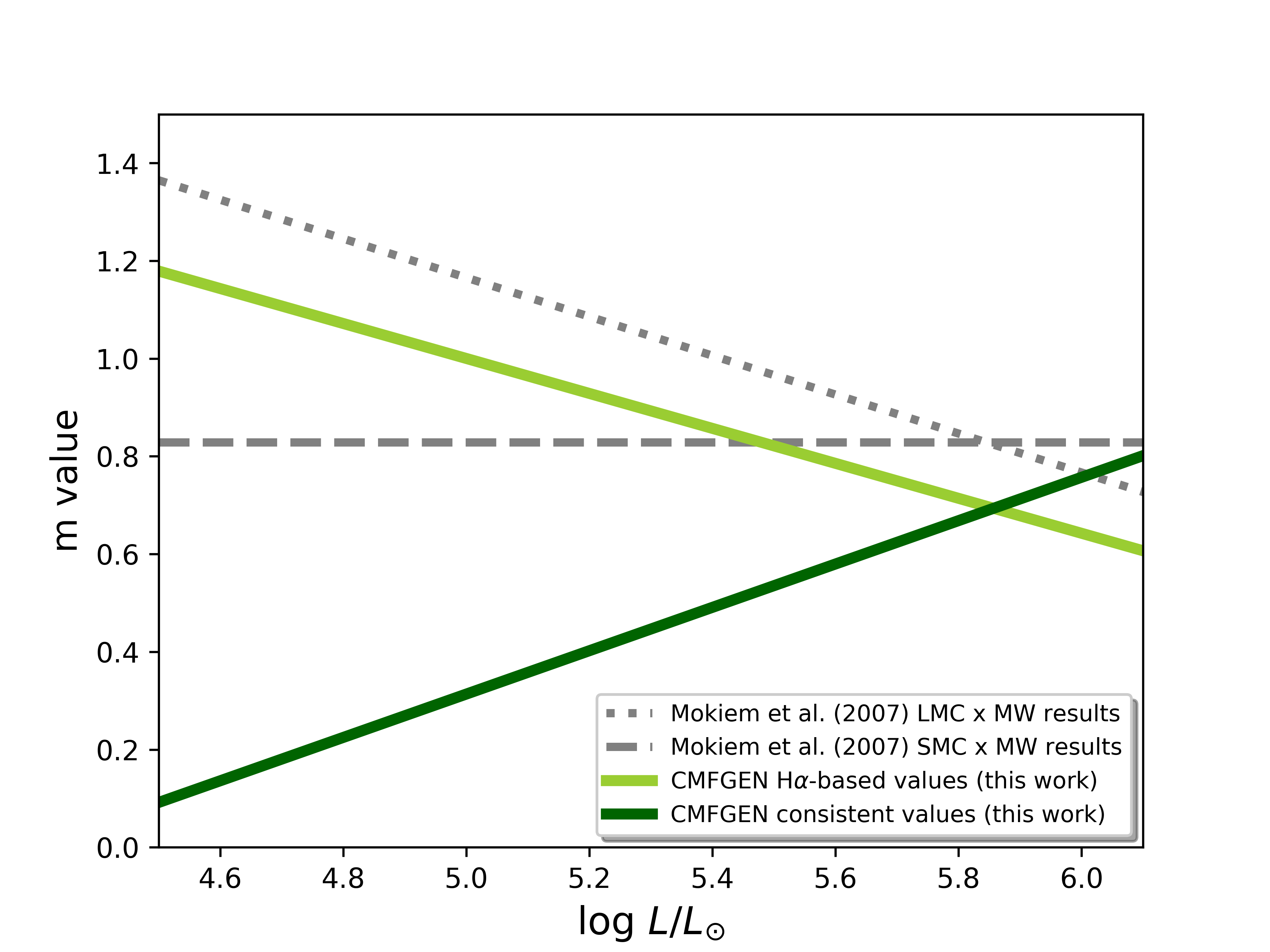}
    \caption{The metallicity dependence ($m$ power) as a function of the luminosity - $\dot{M} \propto $\,Z$^m$.}
    \label{fig:alpha}
\end{figure}

We can now analyse the metallicity dependence from the CMFGEN results presented above. 
We assume that $\dot{M} \propto $\,Z$^m$ and that $v_{\infty} \propto $\,Z$^n$, with n fixed at 0.13, following 
\citet{Leitherer92} (see however our discussion of the terminal velocities below). Therefore, with $\log D = \log (\dot{M}v_\infty\sqrt{R_{\star}})$, we can write the difference: 

\begin{equation}
   \Delta \log D \equiv \log\, D_{MW} - \log\,D_{SMC} = (m+n)\,\log\,\frac{Z_{MW}}{Z_{SMC}}  
\end{equation}

Our fits for the MW and SMC provide (see Fig. \ref{fig:MWSMC} and \ref{fig:Halpha-v1}):

\begin{equation}
 \log\,D_{MW} = \beta_{MW}\,\log\,L/L_{\odot} + \alpha_{MW} 
 \end{equation}

\begin{equation}
 \log\,D_{SMC} = \beta_{SMC}\,\log\,L/L_{\odot} + \alpha_{SMC}     
\end{equation}

Where $\beta$ and $\alpha$ are the angular and linear coefficients, respectively 
(see Table \ref{tab:linearcoeffs}). Combining the equations, we can compute $m$ as a function of the luminosity through:

\begin{equation}
\label{eqform}
m = \frac{ (\beta_{MW}-\beta_{SMC})\,\log\,L/L_\odot + (\alpha_{MW}-\alpha_{SMC}) }{\log\,Z_{MW}/Z_{SMC}} - n 
\end{equation}

In Fig. \ref{fig:alpha} we use Eqn. \ref{eqform} on the empirical results presented in the last sections. We adopt 
Z$_{SMC} = 1/5$ Z$_{MW}$. For comparison, we also display the results by \citet{Mokiem07}, regarding SMC x MW and also LMC x MW. 
With our consistent values for the wind momenta, the $m$ exponent increases with luminosity, a fact which is 
apparent from Fig. \ref{fig:MWSMC} -- where the MW and SMC points start to disentangle. In contrast, a very weak 
Z-dependence is revealed by the fact that $m \rightarrow{0}$ at low $L$. 

If we take into account the H$\alpha$-based D values, we do not observe the same trend. 
At low L, the lines separate even further -- $\dot{M}$(H$\alpha$) > $\dot{M}$(UV), increasing $m$.  
However, the consistent values are the ones to be followed. The H$\alpha$-based values are shown 
here only for completeness (see Sect. \ref{sec:discussion}). 

The $m$ value obtained by \citet{Mokiem07} is 0.83, which is explicit in Fig. \ref{fig:alpha} as an  
horizontal dashed line (SMC x MW). Such a value is only perceived by our results  
at the high luminosity end. At $\log L/L_\odot = 5.75$, the value chosen by \citet{Mokiem07} 
to compute $m$, our results indicate a milder dependence, $m \sim 0.6$. For the LMC x MW, the data explored by 
\citet{Mokiem07} suggest that the Z dependence gets stronger at low L, if extrapolated. But there, 
their results are uncertain.



Overall, the present analysis suggests that the metallicity dependence of O stars is better 
seen at high luminosities. There, the different relations diverge less. 
Regarding our data, for $\log L/L_\odot \gtrsim 5.4$, where we see start to see 
a clear separation of MW and SMC points in Fig. \ref{fig:MWSMC}, the $m$ values fall 
in the interval $0.5-0.8$. Hence, above this threshold, we suggest   
$\dot{M} \propto $\,Z$^{0.5-0.8}$ as a reasonable, realistic relation\footnote{
We remind that we have used Deming regression, which takes 
the errors both in the modified momenta and luminosities properly into account 
when fitting the data. If a simple linear regression is applied, 
$m$ in the interval $0.6-1.2$ is obtained for $\log L/L_\odot \gtrsim 5.4$.}.

\begin{figure*}
	\includegraphics[width=\textwidth]{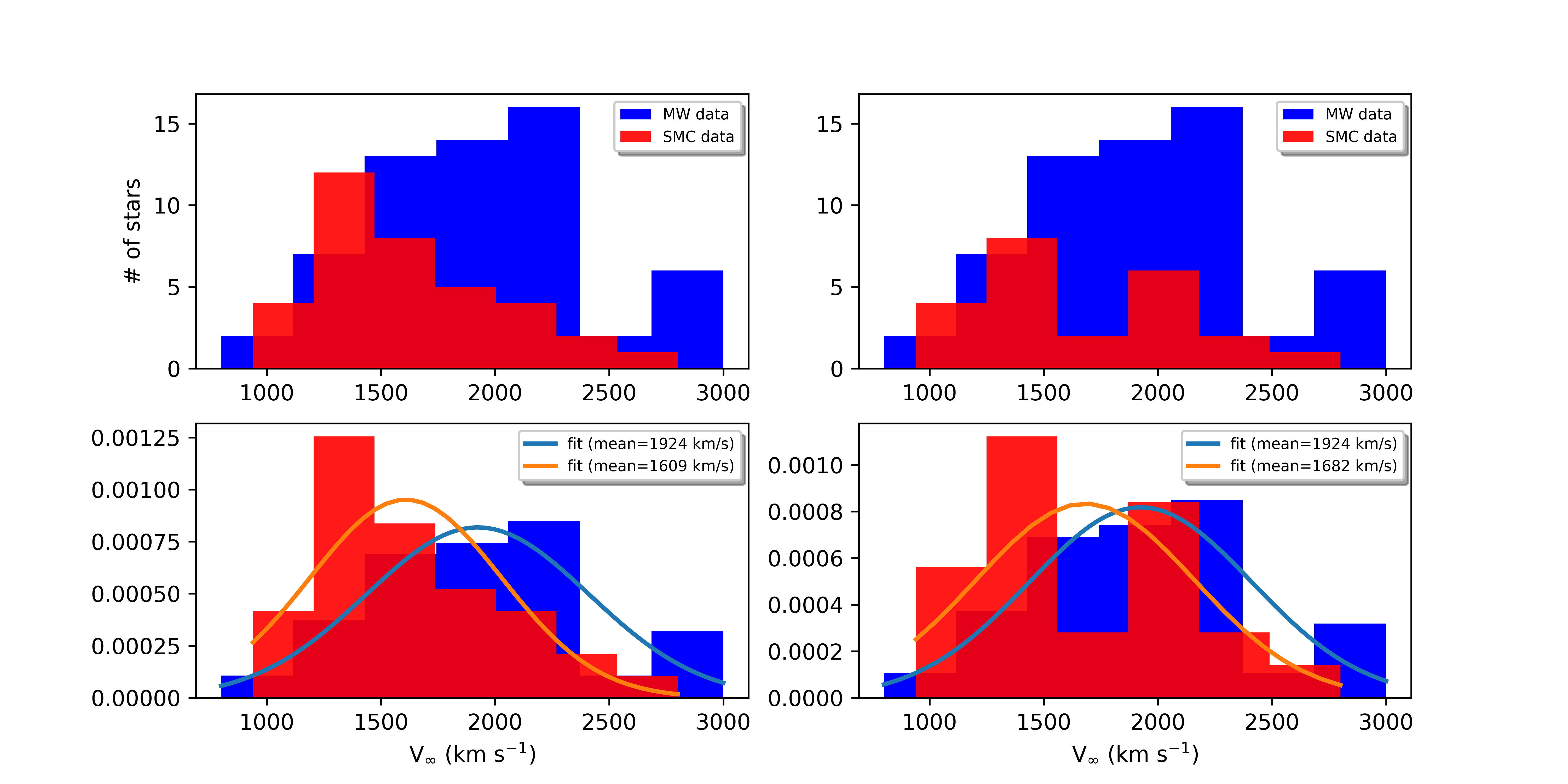}
    \caption{Terminal velocity distribution of our sample of O and B stars in the MW and SMC. Left panels include all our data. Right panels exclude SMC stars without conspicuous wind  profiles in \citet{Bouret13} (see text for more details).}
    \label{fig:vinfauto}
\end{figure*}

On the other hand, the weaker $\dot{M}$(Z) dependence found at low luminosities ($m \rightarrow{0}$) 
deserves to be investigated in future studies, with a larger sample. If such stars indeed loose 
about the same amount of mass per year regardless the environment (Z), 
there are likely evolutionary consequences (e.g., angular momentum, mixing). 
As low $L$ means less massive, this problem would affect the majority of O 
and B stars, as required by an initial mass function (IMF) distribution.

\section{Terminal velocities}
\label{sec:vinfz}

In this section, we analyse the terminal velocities ($V_\infty$) of our sample 
stars to check for a possible metallicity dependence. Some theoretical calculations indicate that $V_\infty$ should scale with metallicity as $V_\infty \propto $\,Z$^n$. 
For example, \citet{Leitherer92} computed a multiple linear regression to their set 
of radiatively driven wind solutions for several theoretical stars in the 
Hertzprung-Russell diagram, of different metallicities, 
obtaining: $\log V_\infty = 1.23 - 0.3\log (L/L_{\odot}) + 0.55 \log (M/M_{\odot}) + 0.64 \log T_{eff} + 
0.13 \log (Z/Z_{\odot})$. The last term implies $n = 0.13$, a value which is widely used in 
the literature. On the other hand, \citet{Krticka18} found from hydrodynamical calculations 
a metallicity dependence for the mass-loss rate ($\dot{M} \propto $\,Z$^{0.59}$), 
but that the average values of $V_{\infty}/V_{esc}$ ($V_{esc}$, the escape velocity) 
for stars in the LMC, SMC and the MW were similar, in contrast to what was found 
by \citet{Leitherer92}. More recently, \citet{Bjork21} inferred a negative but still 
shallow $n$ value (-0.10 $\pm$ 0.18) and \citet{VinkSander21} inferred $n = 0.19$.

Using spectroscopic observations, $V_{\infty}$ can be measured with or without 
atmosphere models. Well developed P-Cygni profiles in the UV can be used for a 
direct determination by using the bluest wavelength of zero intensity (i.e., $V_{black}$) 
or {\it narrow absorption components - NACs} \citep*[see][]{Prinja90}. 
With models, detailed fits to the whole absorption part of the P-Cygni profiles 
provide the measurements, with a depth-dependent turbulence included.
In general, these two methods agree very well within a few percent 
\citep*[see][and references therein]{Crowther06}.

In our sample, most measurements were obtained directly from the comparison 
between observed and synthetic P-Cygni profiles in the UV. In some occasions,  
values from the literature were adopted for $V_\infty$. Nevertheless, the fit quality 
to diagnostic lines is always checked and adjustments are made, when necessary.
For example, \citet{Crowther06} used $V_{black}$ values from previous works 
and \citet{Searle08} used the UV-based results of \citet{Prinja05} as input 
for $V_\infty$, getting reasonable fits to some UV profiles\footnote{We note 
that lines from certain ions were not properly reproduced in their study due 
to absence of X-rays in the models.}.


\citet{Bouret13} achieved satisfactory fits to the wind profiles in 
their SMC sample, obtaining $V_{\infty}$ within $\pm$ 100 km s$^{-1}$. 
However, for stars without or with undeveloped P-Cygni profiles, 
$V_{\infty}$ measurements becomes challenging, 
sometimes impossible. In such cases, \citet{Bouret13} adopted terminal velocities 
of MW stars of similar spectral types from \citet{Kudritzki00} and applied a $V_\infty \propto Z^n$ scaling, 
using $n = 0.13$ \citep*[][]{Leitherer92} and the SMC metallicity. These objects can be identified 
in Table \ref{tab:sampleSMC} when the modified momentum value informed is an upper limit.
Although the observed profiles were well reproduced, the reliability of these $V_\infty$ values is questionable. 
Due to this fact, we will present below an analysis with and without these sources.

\begin{figure}
	\includegraphics[width=\columnwidth]{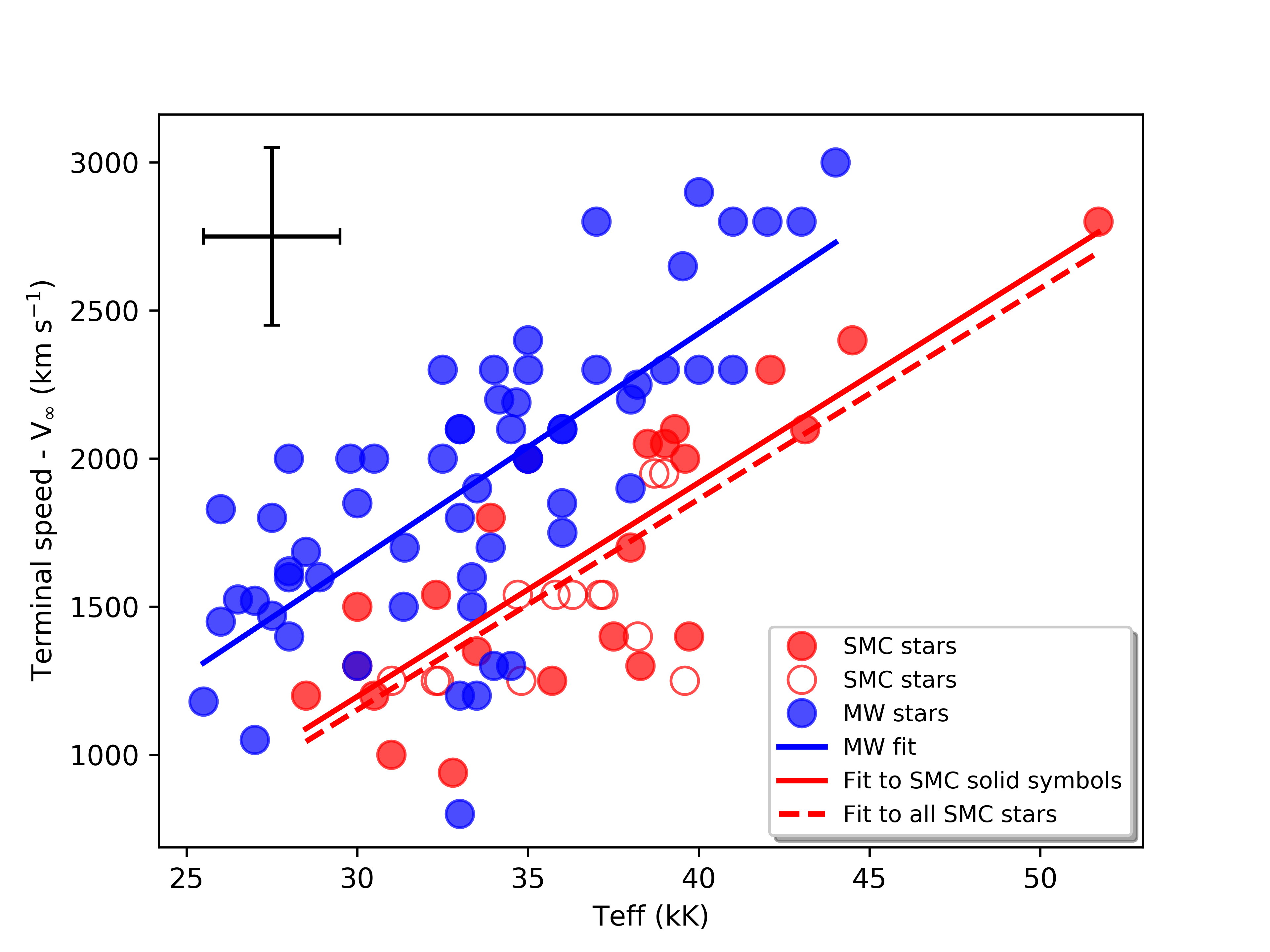}
    \caption{Terminal velocity versus effective temperature for our sample of O and B stars in the MW and SMC. In general, MW stars possess larger velocities if compared with SMC stars. Representative error bars are shown: 2000 K in temperature and 300 km s$^{-1}$ in velocity. SMC stars represented by open circles do not have conspicuous wind profiles (see text for more details). }
    \label{fig:vinfteff}
\end{figure}


We present in Figure \ref{fig:vinfauto} histograms of the terminal velocities listed  
in Tables \ref{tab:sampleMW} and \ref{tab:sampleSMC}. By taking into account all data 
(left panels), a metallicity dependence is suggestive. Indeed, a Kolmogorov-Smirnov test 
to check whether these data come from a same population returns a p-value of 0.002215. 
Normal fits to the two distributions return $\bar{V}_{\infty}(SMC) = 1609$ km s$^{-1}$ ($\sigma _{SMC} = 419$ km s$^{-1}$) and 
$\bar{V}_{\infty}(MW) = 1924$ km s$^{-1}$ ($\sigma _{MW} = 488$ km s$^{-1}$). If we use these mean velocities and assume that $V_\infty \propto $\,Z$^n$, we can roughly estimate that $n \sim 0.11.$ 

Similarly, but neglecting the SMC stars in \citet{Bouret13} that do not have conspicuous wind profiles, 
we obtain that $\bar{V}_{\infty}(SMC) = 1682$ km s$^{-1}$ ($\sigma _{SMC} = 478$ km s$^{-1}$). In 
this case however, we infer $n \sim 0.08$ and the p-value of the Kolmogorov-Smirnov test increases to 0.1348, weakening 
the evidence.




The use of this Z$^n$ scaling is of course valid when we consider the same physical parameters 
(luminosity, mass, temperature) except for the environment metallicity (Z). Despite using mean 
values from stars with different physical parameters, it is interesting that the values obtained 
above are relatively close to the theoretical value inferred by \citet{Leitherer92}, $n = 0.13$. 
The data indicate that the stellar winds of O and B stars in the SMC are roughly $\sim 15$\% weaker 
if compared with the O and B stars in the Milky Way, on average.

A detailed analysis of $V_{\infty}$ taking into account data 
of several O and B stars in the MW, LMC, SMC, M31, M33, and IC 1613, was presented 
by \citet{Garcia14}. It was the first detailed study of terminal velocities that included massive stars beyond the 
Magellanic Clouds. These authors reported a Z dependence, but with IC 1613 stars occupying about the 
same loci than Magellanic Cloud stars in the $V_{\infty}$-$T_{eff}$ diagram 
\citep*[see][for a discussion of the actual metallicity of the IC 1613 galaxy]{Garcia14, Bouret15}.

In Fig. \ref{fig:vinfteff}, we display our data in the $V_{\infty}$-$T_{eff}$ plane, 
as done by \citet{Garcia14}. These authors explored in detail the complexity of the 
$V_{\infty}/V_{esc}$ distribution, showing that it can be very uncertain to estimate 
$V_{\infty}$ from $V_{esc}$ from the canonical 2.65 ratio \citep*[][]{Kudritzki00}.
Here, we refrain from repeating a $V_{\infty}/V_{esc}$ analysis. Instead, we use the linear 
fits in Fig. \ref{fig:vinfteff} to provide a rough estimate for the $n$ power in the 
Z dependence. 

Using that $V_{\infty} = \beta T_{eff} + \alpha$ with the respective coefficients from the MW and SMC 
fits (SMC - all data: $\alpha = -984.1$, $\beta = 71.2$; MW: $\alpha = -646.6$, $\beta = 76.72$; SMC - neglecting 
data: $\alpha = -971.2$, $\beta = 72.2$) and that $V_{\infty} \propto $\,Z$^n$, we can write:

\begin{equation}
  \log \left( \frac{Z_{MW}}{Z_{SMC}}\right)^n = \log \frac{ \beta_{MW} T_{eff} + \alpha_{MW}  }{\beta_{SMC} T_{eff} + \alpha_{SMC}}     
\end{equation}
  
From this expression, considering the complete sample, we infer $n$ values between $0.16-0.22$, 
for the $T_{eff}$ interval $\sim 40-30$ kK, where most points are. Neglecting the same aforementioned SMC stars 
from \citet{Bouret13}, we infer infer a slightly lower range of $n$ values, namely, $0.14-0.20$.

We conclude that the terminal velocity dependence on Z that can be 
inferred from the CMFGEN results -- $V_{\infty} \sim $Z$^n$ -- has $n$ in the range 
$0.08-0.22$, or approximately, $n \sim 0.1-0.2$. A low exponent seems to be appropriate 
for $V_\infty$ scaling in different environments, supporting \citet{Leitherer92}. 

The results above illustrate the need for an analysis of a larger sample of stars. 
In particular, very weak wind profiles should be used with caution and realistic 
error bars should be provided. Moreover, pure absorption lines or doubtful 
wind profiles should be neglected. This would help improve the statistics considerably.

\section{Discussion}
\label{sec:discussion}

We now discuss our findings, caveats, as well as 
other results in the literature.

\subsection{H$\alpha$ uncertainties}

Given the importance of the degeneracy question at low $L$, it is 
appropriate to discuss the mass-loss rates obtained from H$\alpha$ measurements. The 
modified wind momentum-luminosity relation gets more scattered because we favored H$\alpha$ over UV measurements, 
when discrepant, changing the linear fit (see Figure \ref{fig:Halpha-v1}). 

We remind that H$\alpha$-based measurements are not free from problems and sometimes they can 
be very uncertain at mid and low luminosities (giants and specially dwarfs). From an observational 
point of view, H$\alpha$ data is relatively easy to obtain, specially for bright objects (low exposure times), 
where it is in emission, allowing fairly easy mass-loss rate determinations. It is no surprise that 
the first modern quantitative spectroscopic studies in the literature were optical-based, 
favoring OB supergiants\footnote{It is worth noting that the UV and optical diagnostics 
usually provide a single mass-loss rate for these objects.}. Moreover, hydrogen is 
a simpler atom to treat than CNO and Fe-group ions in atmosphere models. These facts provide a 
bias towards H$\alpha$ measurements and bright objects.


However, H$\alpha$ can be mostly in absorption in O giants and dwarfs. The determination of the mass-loss rate  
in such cases is based on the wind emission that fills the core of the photospheric line. 
This can be very uncertain, depending on the object/sample. For example, nebular emission from Balmer lines 
is common around young objects (e.g., in dwarfs). Some cases observed go from weak to strong 
central emission superposed on the H$\alpha$ absorption profile.  
In some other cases, one cannot be sure whether there is a weak nebular or a wind contribution to the 
observed profile, or both. Line-filling from a contamination of a companion star have been also claimed as a 
possibility in the literature. Such difficulties have been reported in different works  
and can hinder a reliable mass-loss rate determination \citep*[see e.g.,][]{Agudelo17, Ramachandran19}. 

Of course, not everything can be assigned to observational issues. There are cases where 
H$\alpha$ seems free of contamination and the models cannot fit it simultaneously with the 
wind diagnostics in the UV. Therefore, it is indeed possible that the models are incomplete 
and inadequate for some objects. Assuming that this is true, on the other hand, 
it is puzzling that for every luminosity class we have examples where models successfully fit  
the observed profiles, from the ultraviolet to optical.

In brief, we have discussed the impact (Sect. \ref{sec:halpharesults}) and uncertainties of the 
H$\alpha$ mass-loss rate measurements. However, {\it consistent rates} - that successfully match the 
UV and optical wind diagnostic lines - are the ones to be taken into account. Obviously, 
the discrepancies found in some objects still deserve investigations. However, until this question 
is settled, there is no reason to give more weight to H$\alpha$ over UV $\dot{M}$ measurements. 

\subsection{UV uncertainties} 

Detailed analyses of UV spectra with CMFGEN provide a way to determine mass-loss rates and 
also terminal velocities, which are necessary to compute the modified momentum 
$D_{mom}$ ($=\dot{M}V_{\infty }\sqrt{R_{\star}}$). However, how robust are these mass-loss 
rates ? In particular, how certain can we be about the possible degeneracy implied by 
Fig. \ref{fig:MWSMC} ?

First, we remind that the rates in Fig. \ref{fig:MWSMC} are consistent. That is, the models present 
a reasonable fit to the UV and optical spectra for the sample stars. However, it is a fact that some 
CMFGEN models fail to fit these two spectral regions simultaneously for a couple of stars, as previously mentioned. 
This raises the question whether CMFGEN really grasp the essential physics present in the winds of 
O stars or provide only rough mass-loss rates in some cases. 


\subsubsection{Clumping}

Radiatively driven winds are far from being homogeneous \citep*[][]{Owocki88}. 
Line instability cause strong shocks within a stellar wind, generating X-rays and a variety 
of effects on the outflow. The winds present time-dependent structures, with strong density 
(clumping), temperature, and velocity variations. In CMFGEN, only an approximate treatment is made 
to incorporate the effects of X-rays and clumping. In general, the clumps are assumed to be optically 
thin (micro-clumping) and the inter-clump medium is void\footnote{The presence of clumping enhances 
recombinations, making $\rho^2$ sensitive lines stronger 
compared with the ones from homogeneous models. Therefore, clumped models mass-loss rates 
must be decreased by a factor of $1/\sqrt{f}$ in comparison with homogeneous models to fit the 
observations \citep*[see][]{Bouret05}.}. 

However, it has been shown that clumps can become easily optically thick in strong UV wind lines. 
The effects of clumps of arbitrary optical depths 
(from thin to thick) and of an inter-clump medium on UV line profiles have 
been explored recently by \citet{Sundqvist18} (see also 
references therein). They incorporated optically thick effects (porosity and vorosity) in the 
FASTWIND code through the use of effective opacities. These authors explored two UV resonance lines, 
from N\,V and P\,V, and found that they get weaker compared with the ones from 
optically thin clump models. This indicates that mass-loss rates from models taking into account 
micro-clumping are underestimated.

In the context of the present work, we remind that: (i) we corrected the mass-loss rates from 
CMFGEN models with (micro-) clumping upwards by a $1/\sqrt{f}$ factor. That is, we used 
only homogeneous values in our plots. For $f = 0.1$ for example, the correction is about a factor of $3$; 
(ii) the changes on the UV line intensities found by \citet{Sundqvist18} from optically thin {\it versus} 
optically thick models are not huge (see their Fig. 4). Although not quoted by these authors, we 
estimate that the mass-loss rate corrections with the optically thick clumping effects included are 
below a factor of 10. Hence, although the WLRs obtained in the present paper can be somehow modified with a 
more realistic description of clumping, we speculate that the impact on the mass-loss rates will be, to some 
extent, compensated by our correction from micro-clumping to homogeneous rates.

The question above will be only settled when a large number of objects 
is analyzed relaxing the micro-clumping assumption, preferably using multi-wavelength diagnostics.
A limitation of the work by \citet{Sundqvist18} is that they explored only UV profiles 
from $\zeta$ Pup like models, i.e., early-type stars. Therefore, we lack a view of the optically 
thick effects on UV lines in different type of objects, e.g., of low luminosities and of different 
metallicities. It remains also to be explored whether the effects will solve the discrepant 
mass-loss rates measurements from the UV and H$\alpha$ made by CMFGEN in a couple of objects 
\citep*[see e.g.,][]{Martins12}.

We note that new alternative approaches to this problem start to emerge. 
\citet{Flores21} for example, represent clumping by dense 
spherical shells with the CMFGEN code. The opacity naturally follows from the 
density variations within the wind. Thus, the interclump medium and arbitrarily 
optical depths are automatically taken into account. 
\citet{Flores21} obtain a mass-loss rate 40\% higher than the micro-clumping 
approach, for the supergiant AzV83 (O7Iaf+). It remains to be seen whether 
this trend will persist for stars of different spectral types and luminosity 
classes. 

Another major concern regarding clumping is its metallicity dependence.
For example, if the filling factors\footnote{One should be careful about the nomenclature.
In our notation, we have the clumping volume filling factor $f \leq 1.0$.
In the literature, the inverse is often called the clumping factor ($D = 1/f$).} 
vary with Z, different mass-loss corrections are expected 
\citep[][]{Sundqvist13}. The WLR for the MW and Magellanic Clouds
stars would then change in a significant way. Currently, this is an open question. 

Interestingly, there are some works in the literature that
favor a Z independent view of clumping. For example, \citet{Marchenko07}
presented time-resolved spectroscopy of 3 SMC Wolf-Rayet stars with the
ESO-VLT-UT2 8-m telescope, finding evidence of small scale structures
moving within the outflow. The measured properties of these clumps -- velocity dispersion, acceleration
and emissivities -- were found to be similar to the ones found in Wolf-Rayet stars of the Milky Way. 
In addition, \citet{LepineMoffat08} monitored optical lines of a small
sample of Galactic O and W-R stars, reporting similar {\it lpv} patterns for all
objects. They concluded that "stochastic wind clumping is a universal phenomenon in
the radiation-driven, hot winds from all massive stars, with similar clumping factors 
in all stages of mass depletion". Hence, the results of both studies -- \citet{Marchenko07}
and \citet{LepineMoffat08} -- naturally lead to the conclusion that clumping is similar in stars of
different evolutionary stages, regardless the environment.  Unfortunately, we lack similar 
studies confirming this view, for a larger sample of stars.

It is also worth noting that there is no indication that the filling factors and the onset of clumping
are drastically different in MW and Magellanic Cloud objects, with the usage of 
micro-clumping in CMFGEN models. If there is a Z dependence, it 
is subtle to the point of not being perceived by such standard analyses.

\subsubsection{Weak winds}

Mass-loss rates inferred for low luminosity O stars were found to be
much lower than predicted by theory \citep*[e.g.,][]{Martins05, Marcolino09, Puls08}.
For example, UV line profiles obtained from atmosphere models match the observed ones in O8-9V stars when mass-loss 
rates are typically $10^{-9}$ M$_\odot$ yr$^{-1}$. For these same type of objects, theory predicts roughly 
$10^{-7}$ M$_\odot$ yr$^{-1}$. This problem challenges our understanding of the winds of these objects.
In the context of the present work, if there is an issue with the atmospheric measurements
and the winds are not weak -- e.g., they are as predicted by theory \citep*[][]{Vink01} -- our results at low L
are called into question.

In the last decade, several works in the literature approached the weak wind problem.
\citet{Lucy10}, using a Monte Carlo technique, obtained theoretical mass fluxes somewhat compatible with the ones observed 
in late-type O stars, about 1.4 dex lower than \citet{Vink01}. On the other hand, by updating the work of 
\citet{Vink01}, \citet{Muijres12} reported that their models failed to produce winds at low
luminosities due to a lack of Fe V lines, which possibly indicates the need of other mechanisms to help the wind driving.

\citet{Vilhu19} suggested that the solution is related to the velocity span within a wind clump
(using a velocity filling factor FVEL = 0.1). When taken into account, the radiative force
was modified and the mass-loss rates observed for late-type O stars (obtained by atmosphere models) were matched by
their models. Notwithstanding, the physical reason for the specific value of the velocity filling 
factor used remained unclear. Also unclear was the fact that this effect was not needed for brighter objects (FVEL = 1). 

\citet{Sundqvist19} provided mass-loss rates from hydrodynamical simulations that used 
a radiative force computed from co-moving NLTE radiative transfer solutions, in 
a self consistent way. For parameters close to that of an O7V star, the rate obtained was about 9 times 
lower than predicted by \citet{Vink01}, a tendency towards weak winds but not a solution. 
New simulations by \citet{VinkSander21} and \citet{Bjork21}, which we discuss later in the paper, 
maintain relatively high mass-loss rates for a typical late O type star and therefore the weak wind problem.

An interesting solution was recently proposed by \citet{Lagae21}, which carried out hydrodynamical simulations that
took into account the line-deshadowing instability (LDI). The LDI naturally creates a very structured wind, with
drastic variations in velocity, density and temperature. For low luminosity O stars, their simulations show that 
most of the outflowing gas is shocked, with temperatures well above 10$^5$ K. The line profiles computed with these
models were weaker than the ones from homogeneous models, for a same mass-loss rate. This indicates that the
mass-loss rates are underestimated if the shocked region is not taken into account. That is, all current UV based rates 
from atmosphere models are questioned at low luminosities.

In order to confirm this scenario, the calculation of the amount of x-rays emitted from their models
and a comparison with the observations is necessary. If about $\sim 70\%$ of the wind volume of an O dwarf has 
temperatures about $10^5-10^7$ K, as found by these authors, a large contribution to the 
integrated x-ray emissivity and thus total x-rays emission, is expected. Similarly, we expect 
less x-rays emission from their O supergiant model, as the volume fraction of the 
shocked, hot wind is only about $7\%$ and on average its temperature is below 10$^5$K. 
A priori, x-rays observations seem to be in contrast with these considerations 
unless the different gas densities in these two cases counteract the very different volumes and temperatures 
of the hot gas. It is known that for O stars log $L_X/L_{BOL} \sim -7.0$ \citep*[see e.g.,][]{Sana06, Oskinova06}. 
About the same value is observed in O dwarfs and supergiants. High (low) luminosity stars having high (low) $L_X$. 

In this same context, we note that some X-rays results support the reality of weak winds, 
with mass-loss rates as low as the UV measurements \citep*[e.g.,][]{Cohen14, Doyle17}. Interestingly, the work of \citet{Huene12} presents an X-ray based mass-loss rate of about $2.0 \times 10^{-9}$ M$_\odot$ yr$^{-1}$ for the O9.5V star $\mu$ Col. Although 
the authors claim that the wind is not weak, this value is not far from the ones found for other 
late-type O stars \citep*[see e.g., $\zeta$ Oph in][]{Marcolino09}. Also, 
the value predicted from theory (Vink) for this star is $1.1 \times 
10^{-8}$ M$_\odot$ yr$^{-1}$, which means a considerable difference compared with 
the observed value (of about 0.8 dex).

Despite several attempts, there is not yet a solution for the weak wind problem.
Some of the works mentioned above decrease the mass-loss rates in comparison 
with Vink et al. and even match the very low rates measured by atmosphere models, but 
there is no consensus so far. In any case, there are no clear evidences to disregard atmosphere model results at low L, as the ones presented here. Some theoretical calculations have shown that very low rates are plausible.
    
\subsection{Other measurements and techniques in the literature}

Recently, \citet{Ramachandran19} presented an analysis of 320 stars 
at the SMC: 297 B stars and 23 O stars. The authors used the 
Potsdam Wolf-Rayet code (PoWR) to model their spectra and derive stellar and 
wind parameters. For the majority of the sample, $\dot{M}$ was 
obtained as an upper limit from H$\alpha$. Given the location of the 
observed stars in the SMC - the supergiant shell SMC-SGS1 - severe 
ISM contamination remained in this line even after the sky subtractions performed. 
This is an example of how uncertain H$\alpha$ can be in some cases. We 
do not consider their H$\alpha$ measurements here.

For nine objects in their sample, low resolution (6-7\AA) {\it IUE}  spectra are available 
and were used. One supergiant - identified as SMCSGS-FS 310 - also has a {\it HST} 
({\it Hubble Space Telescope}) high-resolution spectrum (resolving power R=18000). 
The uncertainties reported by Ramachandran et al. for the mass-loss rates and 
terminal velocities are relatively small, $\pm$0.2 dex and about 10\%, respectively. 
We remind that all UV results from CMFGEN used in the previous sections are based 
on high-resolution data.

\begin{figure}
	\includegraphics[width=\columnwidth]{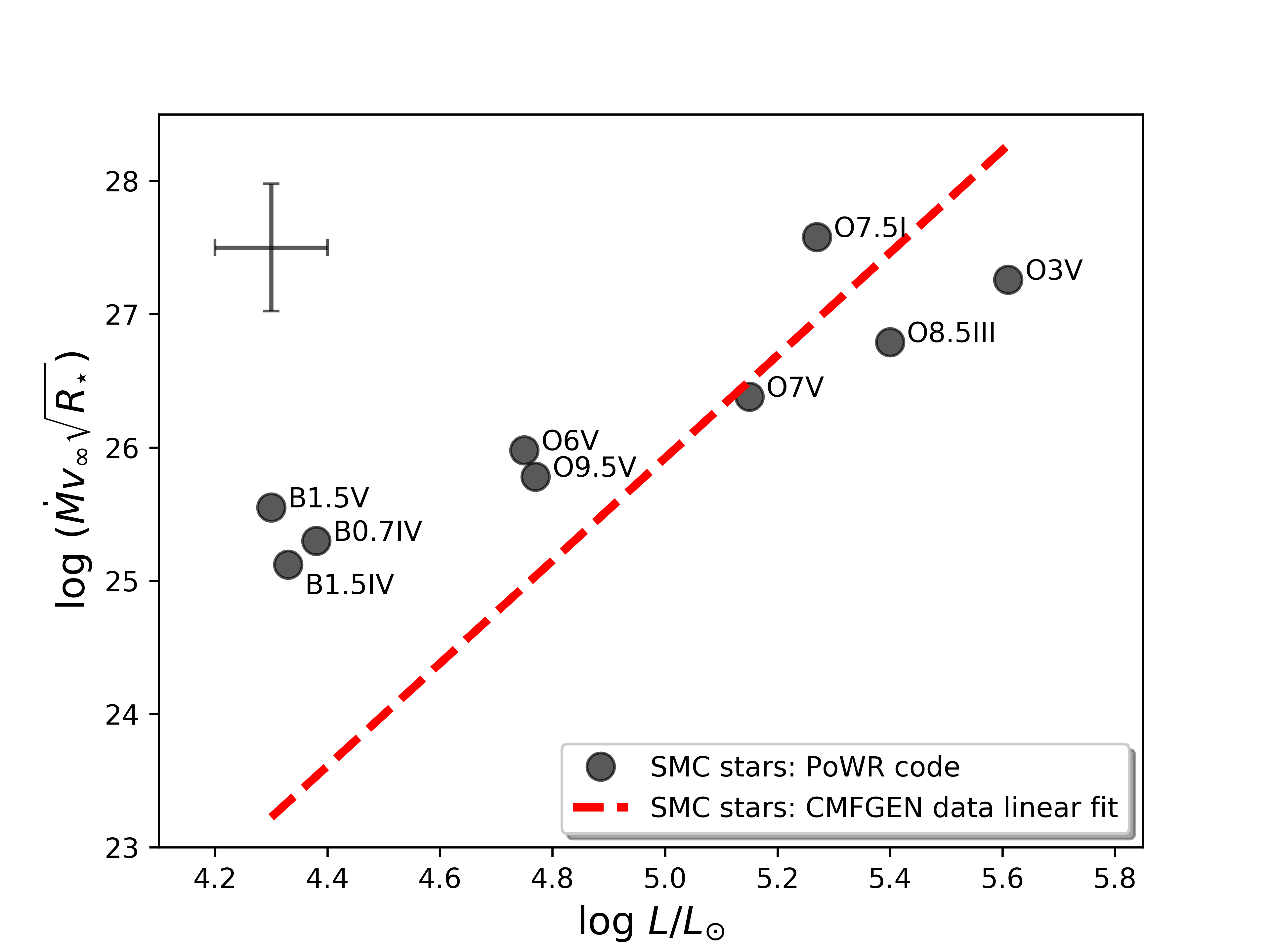}
    \caption{WLR diagram for 9 SMC stars analyzed with the PoWR code. The linear fit is based on our CMFGEN data for the SMC stars. Representative error bars and spectral types are indicated. Note the good agreement for the O stars and that B stars stand above the fit (see text).}
    \label{fig:ramanchandran}
\end{figure}

In Fig. \ref{fig:ramanchandran}, we present the WLR diagram of the nine stars analyzed 
with the PoWR atmosphere code by \citet{Ramachandran19}, the ones with UV data, 
along with the linear fit for our (CMFGEN) SMC sample (see Fig. \ref{fig:MWSMC} and Table \ref{tab:linearcoeffs}), 
extrapolated down to $\log L/L_\odot \sim 4.3$. Nominal values for the mass-loss 
rates from their Table B.2 are used, since they are independent of the adopted clumping 
parameters \citep*[see][]{Ramachandran19}. 

The B stars, which are cooler than all stars considered in our CMFGEN 
sample (at low $L$), stand above the fit. A possible explanation for 
this behavior is bi-stability. The temperatures of 
the B1.5V, B1.5IV, and B0.7IV stars are 22kK, 20kK and 27kK, respectively 
\citep{Ramachandran19}. From about 27.5kK to 22.5kK, the mass-loss rates  
and therefore $\log D$, are expected to increase by some factors due to the 
bi-stability phenomenon \citep{Vink01}. 

On the other hand, there is good agreement for the O stars. This fact supports 
the physical parameters obtained with CMFGEN and the metallicity trend 
discussed in Section \ref{sec:mdotz}. We remind however that the 
determination of accurate stellar wind parameters for metal poor 
late-type O (or early B) dwarfs from low resolution UV spectra is challenging. 
Both characteristics, spectral type and low $Z$ environment, combine to produce 
wind profiles that are very weak to be seen even in high resolution data. 
In this context, it would be reassuring to have a PoWR re-analysis of 
Ramachandran's sample based on new high-resolution spectroscopic observations.

Regarding the Large Magellanic Cloud (LMC), the VLT-FLAMES TARANTULA survey resulted in several 
papers on the physical and chemical properties of O and B stars. Here, we briefly discuss  
the papers by \citet{Agudelo17} and \citet{Sabin17}, which addressed stellar wind parameters. 
Both used optical data.

\citet{Agudelo17} analyzed a sample of 72 giants, bright giants and supergiants massive stars 
at the LMC with an automatic fit procedure using FASTWIND (PIKAIA). About 40\% of their sample, 
the low $L$ stars, have only upper limits on $\dot{M}$. These stars were neglected in their WLR analysis and
linear fits were provided only considering stars with $\log L/L_{\odot}>5.0$, in contrast to our results.
The fits are close to the ones of \citet{Mokiem07} for the LMC. 

Regarding the population of dwarfs in the LMC, \citet{Sabin17}  analyzed 105 O and B stars, 
also through a robust automatic fit method (IACOB-GBAT). Given 
the low metallicity content and luminosity of dwarfs, again only upper limits on mass-loss rates could 
be provided for the latest spectral type stars ($\log L/L_\odot \leq 5.1$). 
Again, the linear fits were provided considering only bright objects, $\log L/L_{\odot}>5.1$.

Although we do not address LMC stars in the present paper, we remind that our 
fits are steeper for the SMC and MW in comparison with Mokiem's relations. 
We anticipate that an analysis of LMC stars data from CMFGEN models will be just 
in marginal agreement with Mokiem's, \citet{Agudelo17}, and \citet{Sabin17} results, at 
high luminosities.

In these works, the terminal velocities are not measured directly. In the absence of UV data, 
they are usually estimated from the escape velocity through the relation 
$v_\infty/v_{esc} \sim 2.6$, scaled to the LMC or SMC metallicity. As discussed before,  
\citet{Garcia14} show that the use of this specific value may be prone to 
substantial errors. This is a source of uncertainty for the (already) upper limits on $\log D$ 
of \citet{Sabin17} and \citet{Agudelo17} for the low $L$ stars.

In short, optical-based only measurements of mass-loss rates of low luminosity O and B stars in Magellanic 
Clouds should be considered with caution. The $\log D \times \log L/L_\odot$ trend at low $L$'s  
is better addressed with our results \citep*[and of][]{Ramachandran19}, based on fits to UV and optical  spectra. 

%
%




\subsubsection{Bow-Shock measurements}

Recently, a very promising technique based on bow shocks to infer 
$\dot{M}$ has been explored in the literature \citep*[e.g.][]{Gvaramadze12, Henney19, Kobul19}. 
The winds of massive stars can produce 
shocks when encountering the ISM medium, producing arc-shaped emission that can be 
successfully detected and imaged at infrared wavelengths (e.g., at 24$\mu$m). 
The mass-loss rate can be expressed in terms of the physical 
properties of this shocked region (e.g., dust emission coefficient, infrared surface brightness). 
This technique is considerably diverse than quantitative multi-wavelength spectroscopy, since 
no comparison with observed spectra or other observable is done. Nevertheless, it is of extreme 
importance as an independent check of the mass-loss rates obtained so far from other techniques.

We focus here on the latest work by \citet{Kobul19}, given the large sample analyzed 
homogeneously. These authors obtained 
mass-loss rates for 67 galactic O and B stars with bow-shock nebulae. The results were 
compared with the predictions of \citet{Vink01} in the $\log \dot{M} \times \log T_{eff}$ and 
WLR diagrams. Overall, an average discrepancy with the predicted 
values (Vink) of -0.43 dex was inferred for stars with $T_{eff} > 25$ kK, with a large dispersion 
of 0.64 dex (see their Fig. 9). Regarding the WLR diagram, a systematic offset of about 0.4 dex was 
also observed, below the predicted relation. That is, in general, 
the mass-loss rates or wind strengths derived were less than predicted by the radiatively driven 
wind theory, which is followed closely by the relations of \citet{Mokiem07}.

\begin{figure}
	\includegraphics[width=\columnwidth]{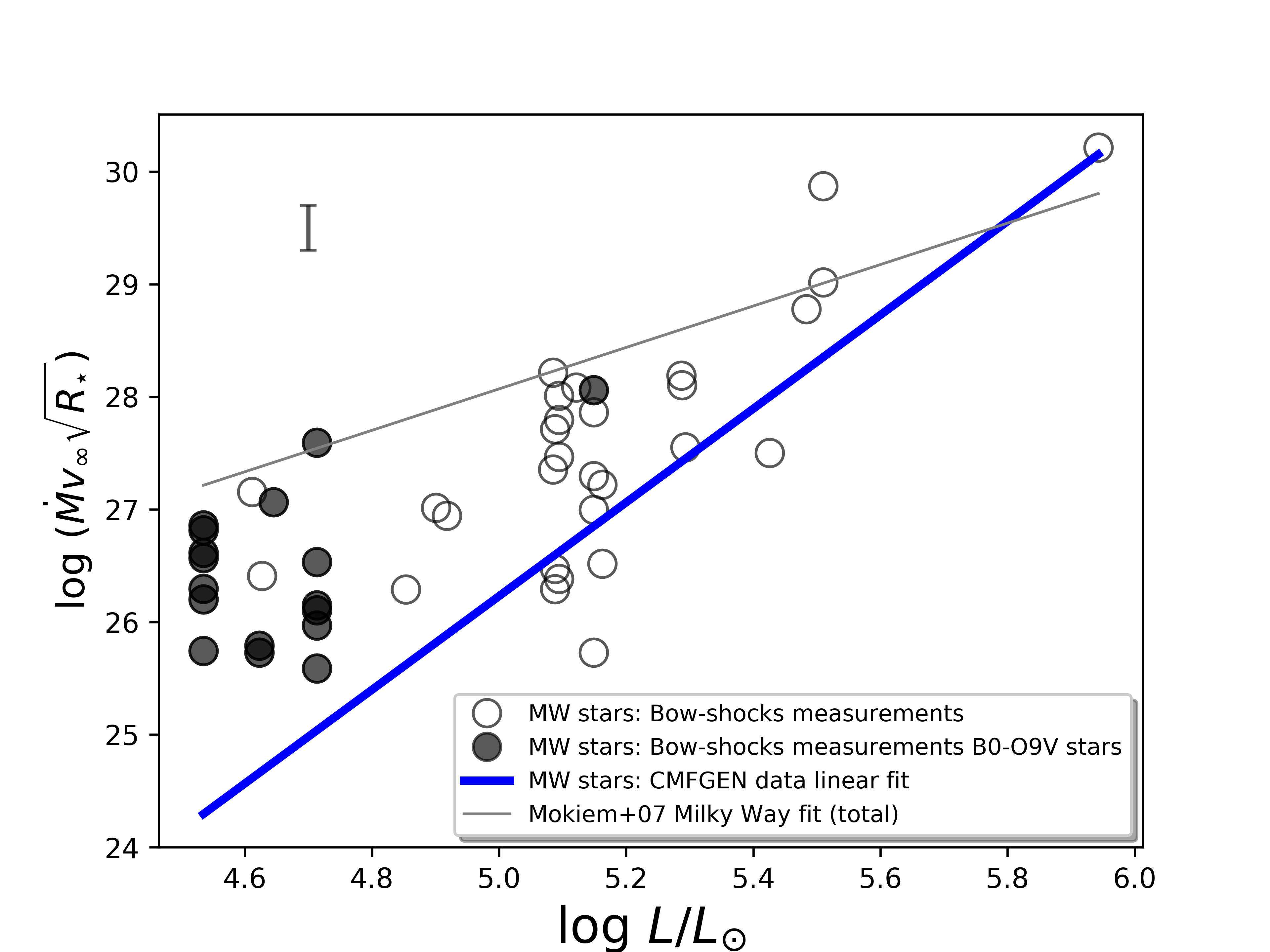}
    \caption{WLR diagram for Milky Way stars, adding measurements based on bow-shocks by \citet{Kobul19}. 
    Linear relations from the results obtained with CMFGEN and from Mokiem's work 
    are displayed (see Section \ref{sec:mdotz}). B0-09V stars are shown with filled circles (see text).}
    \label{fig:WLRkob}
\end{figure}

We present in Fig. \ref{fig:WLRkob} the linear fits obtained from CMFGEN and Mokiem's MW data, 
along with the bow-shock measurements of \citet{Kobul19}. The error bar is representative and 
assumes an uncertainty of 40\% for the mass-loss rate, as estimated by \citet{Kobul19}, neglecting 
errors on the wind terminal velocity and stellar radius. To avoid low effective temperatures ($T_{eff} \lesssim 25kK$) 
and thus the onset of bi-stability, only O and B stars with spectral types earlier than B0 are displayed. 

\begin{figure}
	\includegraphics[width=\columnwidth]{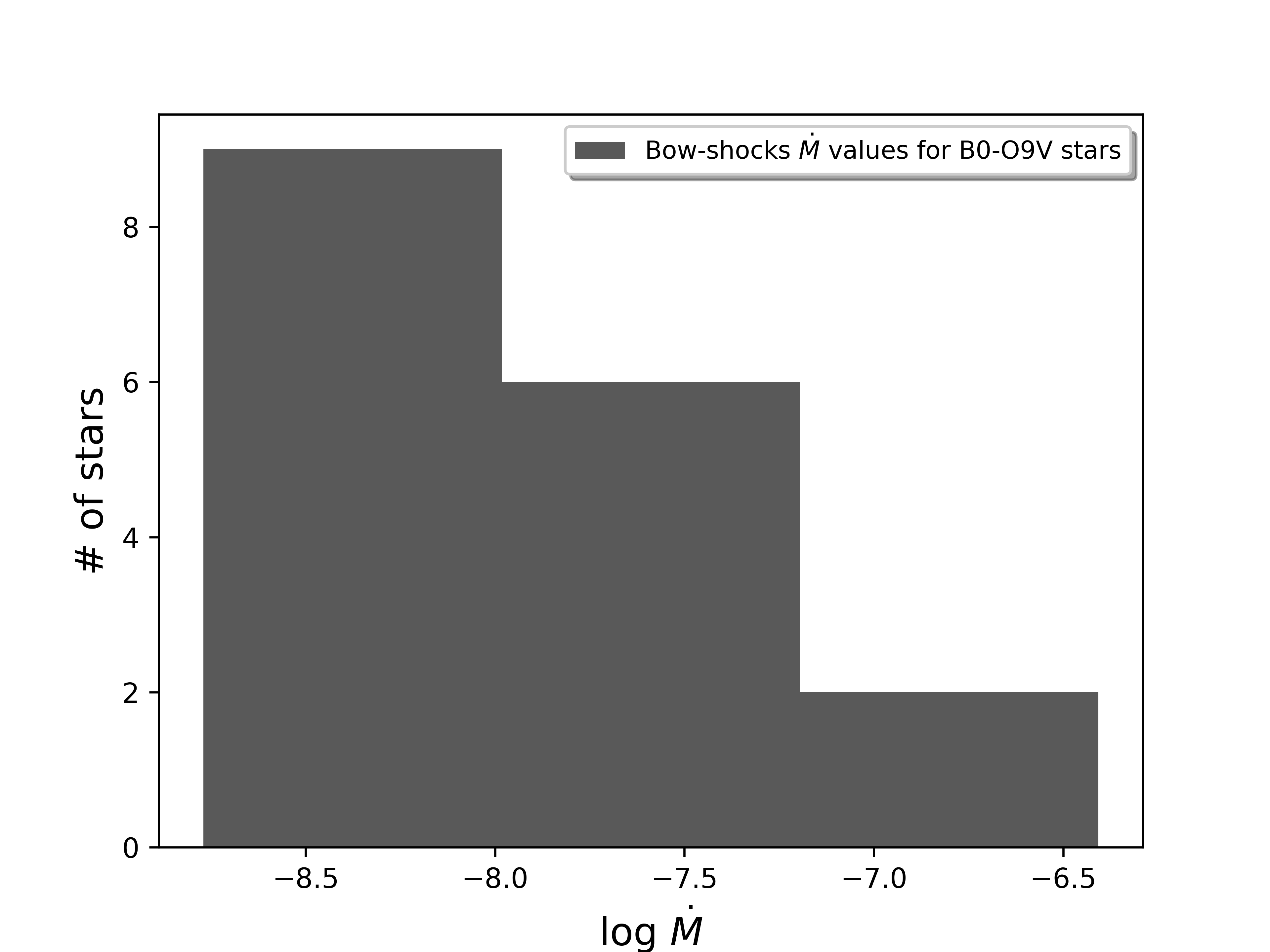}
    \caption{Histogram of mass-loss rates of B0-O9V stars from bow shocks measurements \citep{Kobul19}.}
    \label{fig:histkob}
\end{figure}

Most points fall between the CMFGEN and Mokiem's relations. There is marginal agreement with the 
results from CMFGEN for only about 12 objects out of 46 (about 25\% of their sample). 
These measurements suggest a less steep WLR and the removal of the degeneracy mentioned  
in Section \ref{sec:mdotz}. However, there are some points that deserve to be discussed.

First, we note that there is a considerable scatter for specific 
luminosity ranges/spectral classes in Kobulnicky et al. results. Regarding late O or early B dwarfs (O9-B0V), 
important in the context of the present study, they found mass-loss rate 
values ranging from $\sim 10^{-9}$ to $\sim 10^{-7}$ 
$M_\odot$ yr$^{-1}$ (see their Fig. 11 and Table 3). This can be seen in Fig. \ref{fig:WLRkob}, 
where the B0-O9V stars for example - emphasized as filled circles - span about 2 dex in wind strength, $\log D$. 

To better illustrate this, we present in Fig. \ref{fig:histkob} the range of mass-loss rate 
values for the B0-O9V stars of Kobulnicky et al. work. A priori, a considerable range in mass-loss 
should be supported by very different UV spectral characteristics of these 
stars. However, the observations indicate otherwise, i.e., a fairly 
uniform UV morphology, with similar weak/very weak P-Cygni profiles at these spectral types 
\citep*[see e.g.,][]{Walborn85}. Therefore, we argue here that we should 
take the bow shocks results with some caution, at least when the scatter is high for 
specific spectral classes. Note that CMFGEN results for O9-O9.5V stars have $\dot{M} \sim 10^{-9}$ M$_{\odot}$ 
yr$^{-1}$ (see Table \ref{tab:sampleMW})\footnote{One exception is HD~46202, where a single atmosphere model 
cannot fit the UV and optical data. This star has two measurements for the mass-loss rate. The UV based is in 
fact $\sim 10^{-9}$ M$_{\odot}$ yr$^{-1}$. The H$\alpha$ based is $\sim 10^{-7}$ M$_{\odot}$ yr$^{-1}$, 
a discrepancy so far unsolved}.

The bow-shock method is not free from uncertainties. 
Through Monte Carlo calculations, the average uncertainty for the mass-loss 
rate estimated by these authors is about 40\%, neglecting dust emission coefficients and stellar peculiar 
velocities. In particular, dust emission coefficients can be changed in different ways (e.g., grain destruction 
by shocks), thus affecting $\dot{M}$, as discussed by \citet{Kobul19}. Hopefully, future 
works will help estimate the magnitudes of these uncertainties. 

In conclusion, bow-shocks measurement fall mostly below the predictions of Mokiem's relation 
and above the measurements made with CMFGEN. A reconciliation or at least 
the identification of the physical origin behind the discrepancy between both results 
(CMFGEN and bow-shocks) should be investigated in future studies.


\subsection{Comparison with theoretical predictions}
\label{sec:theory}

The main objective of the present paper is the analysis of empirical data 
of the winds of O and B stars in the MW and SMC. Nevertheless, it is useful 
to compare our results also with theoretical predictions in the literature to check 
for trends and highlight possible problems. We will focus on the recent works 
by \citet{Bjork21} and \citet{VinkSander21}. 

\citet{Bjork21} predict mass-loss rates and terminal velocities for O-type stars 
in the MW and in the Magellanic Clouds, self-consistently. To solve the hydrodynamics, 
co-moving NLTE radiative transfer models are used to obtain the radiative force, in an 
iterative scheme. By comparing the three environments (SMC, LMC and MW), the mass-loss rate dependence 
suggested was $\dot{M}$(Z) $\sim Z^{0.95}$ (i.e., $m = 0.95$), which is a mean. 
The value of $m$ was observed to vary across the luminosity range, approximately 
as $m(L) = -0.32 \times (\log L/L_\odot - 6.0) + 0.79$. Roughly, it goes from about 1.2 at low 
L to 0.8 at high L, in contrast to our data tendency (see Fig. \ref{fig:alpha}).

These authors compare their log $\dot{M}$ $\times$ log $L/L_{\odot}$ predictions 
for MW stars to others in the literature, namely, from \citet{Vink01}, \citet{Lucy10} and \citet{KK17}. 
There is a considerable spread of $\dot{M}$ values for log $L/L_\odot < 5.2$ 
(about $\sim$1 dex; see their figure 5). That is, theoretical predictions do not 
agree at low L. 

Regarding the observations, their theoretical WLR match well the observations 
of bright objects in the MW, that is, for log $L/L_\odot > 5.2$.  
For the SMC, their log $\dot{M}$ $\times$ log $L/L_{\odot}$ predictions are in good agreement 
with the empirical values of \citet{Bouret13}, which we use in this paper
\footnote{Note that \citet{Bjork21} mixed clumped and unclumped mass-loss rates data 
from the table 2 of \citet{Bouret13} in their comparison. Nevertheless, their 
fit is still reasonable when clumping corrections are taken into account, i.e., 
making clumped rates unclumped through $\dot{M}/\sqrt{f}$ = constant.}. By consequence, 
their WLR (SMC) is somewhat close to the one we obtain (see below).

In Fig. \ref{fig:WLRtheory}, we present the WLR obtained in this paper and 
from \citet{Bjork21}. We use their equation 19, for the MW and SMC metallicities. 
The physical units used in their work are modified to conform with ours. 
The expected proportionality between the wind strength and the luminosity, as well 
on metallicity, is apparent. However, a relatively stronger (weaker) Z dependence at low (high) L can be seen in the 
theoretical predictions -- a fact that is emphasized by the authors --  whereas 
our empirical data show otherwise (see also Fig. \ref{fig:alpha}). 
At high L, their WLRs are below the ones inferred by us. The empirical data at low L 
naturally make the slopes of our fits higher. We provide 
an explanation for these differences behaviors below.

\begin{figure}
	\includegraphics[width=\columnwidth]{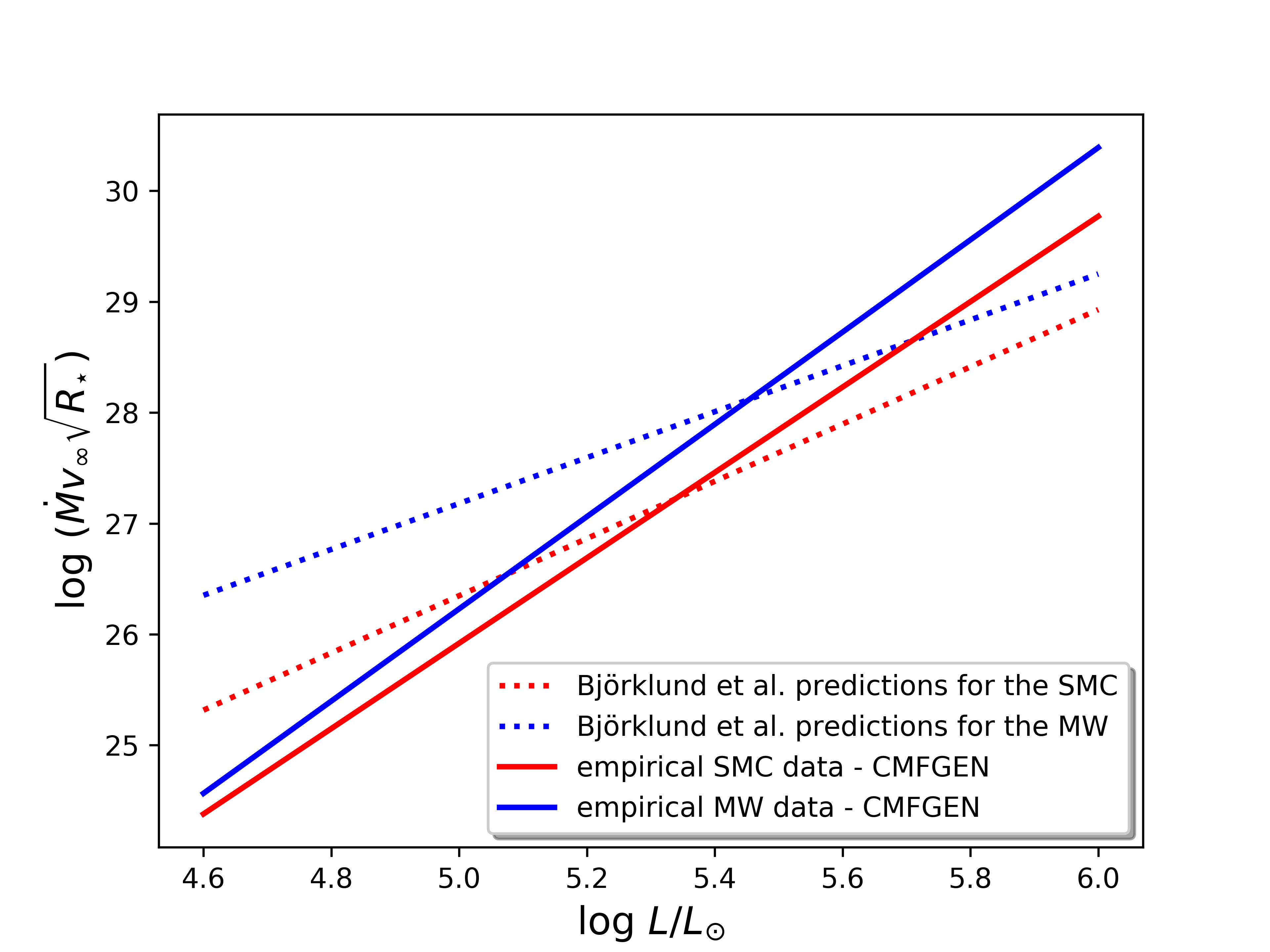}
    \caption{WLR comparison: theoretical calculations from \citet{Bjork21} and our empirical (CMFGEN) data.}
    \label{fig:WLRtheory}
\end{figure}

The predictions of \citet{Bjork21} for MW low L stars -- log $L/L_{\odot}$ < 5.0 -- are 
mass-loss rates from -8.28 to -8.04 (log $\dot{M}$ units). 
The average is -8.14. This is higher than found from  
atmosphere models for O8-9.5V stars by up to an order of magnitude 
(see Table \ref{tab:sampleMW}). For SMC stars, these authors predict 
at low L mass-loss rates from -9.27 to -8.76 (log $\dot{M}$ units). 
The average is -9.06. In contrast with the MW case, this is more in line with 
the rates obtained from atmosphere models (see Table \ref{tab:sampleSMC}).

Therefore, at least part of the discrepancy with our results is related to the weak 
wind problem, which was discussed in Section \ref{sec:discussion}. In short, 
hydrodynamics calculations indicate $\dot{M}$ values that are higher 
than inferred from observations, impacting the WLR -- $D_{mom} \propto \dot{M}$.
For their MW models, the WLR slope is 2.07 $\pm$ 0.32. From our models, we get a much 
higher value, of 4.16 $\pm$ 0.23 (see Table \ref{tab:linearcoeffs}). 
Regarding the SMC, our results indicate 3.85 $\pm$ 0.29 and 
\citet{Bjork21} 2.56 $\pm$ 0.44, which is not a drastic difference as in the MW case.

In addition, we note that the terminal velocities of the low L models 
of \citet{Bjork21} are very high. For example, their log $L/L_{\odot}$ = 4.91 
(T$_{eff}$ = 33383 K) model for a MW metallicity results in $V_\infty$ = 5411.75 km s$^{-1}$. 
Values higher than about 3000 km s$^{-1}$ are rarely reported from observations 
and deserve a deeper investigation \citep*[see e.g., Table \ref{tab:sampleMW} and][]{Prinja90}. 
Such extreme velocities also help explain why their inferred relations are above 
ours -- $D_{mom} \propto V_{\infty}$. In fact, for MW stars with photospheric parameters 
close to the ones aforementioned, some observed $V_{\infty}$ values 
can be as low as $\sim 1000-2000$ km s$^{-1}$. This represents a factor of $\sim 2.7-5.0$ of 
difference, which translates to $\sim 0.4-0.7$ dex in log $D_{mom}$.

\citet{Bjork21} argued that such extreme velocities would not be seen 
in the UV wind diagnostics of low L objects. That is, using UV P-Cygni profiles 
to obtain $V_{\infty}$ would be misleading as most of the wind is predicted to be shocked and at very 
high temperatures \citep*[][]{Lagae21}. However, we note that some terminal velocities obtained 
from x-rays measurements are not very high. On the contrary, they seem to agree 
relatively well with the UV-based terminal velocities. For example, the late-type star 
$\zeta$ Oph has $V_{\infty}$(UV) $\sim$ 1500 km s$^{-1}$ and $V_{\infty}$ (x-rays) $\sim$ 1400 km s$^{-1}$ 
 \citep*[][]{Cohen14}. $\mu$ Col (O9.5V) is another example. \citet{Huene12} report 
 $V_{\infty}$(x-rays) $\sim$ 1600 km s$^{-1}$, which agrees with the UV measurements within 
 error bars\footnote{Note that \citet{Huene12} report that a good fit to 
\ion{O}{viii} (18.967, 18.973\AA) can be also achieved with a terminal velocity of 2800 km s$^{-1}$, 
by increasing the value of the $\beta$ parameter. Despite higher than the UV, this value is still much lower 
than reported by hydrodynamical predictions.} \citep*[][]{Martins05}.
Moreover, as discussed in Section \ref{sec:discussion}, at a first glance, a very large (small) volume 
of hot gas seems incompatible with the x-rays emission observed in these low L (high L) massive stars. 

Very fast winds are also reported in the predictions made by \citet{VinkSander21}. 
Their dynamically consistent computations for the 
mass-loss rate and terminal velocity of massive stars update their previous 
work \citep*[][]{Vink01}. For specific stellar masses and luminosity pairs, 
several Monte Carlo models were computed for different 
metallicities (from 1/33 to 3 Z$_{\odot}$) and temperatures. 

Given the importance of the high terminal velocities found, \citet{VinkSander21} computed 
synthetic line profiles for specific models using the PoWR atmosphere code. 
In summary, they found strong absorption troughs in \ion{C}{iv} $\lambda$ 1549, 
exceeding 5000 km s$^{-1}$. However, it was shown that these 
troughs can be easily masked by carbon depletion at the surface, thanks to CNO mixing. 
Despite this effort, a deeper investigation is needed to address this issue. 
For example, stars close to the ZAMS are not expected to have strong carbon depletion 
and a priori could reveal such very fast wind signatures. A careful comparison with 
observed data is needed.

Regarding the metallicity dependence, for stars above the bi-stability 
jump -- i.e., O stars range -- \citet{VinkSander21} found that $\dot{M}$(Z) $\sim$ Z$^{0.42}$, 
which is much weaker than inferred by \citet{Bjork21}. Their results also indicated 
that the modified momentum depends not only on L, but on T$_{eff}$ as well, possibly hindering 
analyses as done here and in \citet{Mokiem07}.

The T$_{eff}$ dependence of $D_{mom}$ was observed by these authors exploring 
two temperature values, namely, 20 and 40 kK, over a luminosity
interval from log $L/L_{\odot}$ $\sim$ 5.0 to 6.0 (see their Fig. 13). 
In brief, the wind momentum was found to be different for two stars with a same luminosity, if they 
possess different temperatures. Although the results are solid from a theoretical point of view,
we note that there are no observed O stars with T$_{eff}$ as low as 20kK.
They all possess temperatures above $\sim$28 kK \citep*[see e.g.][]{Martins05}.
Similarly, for example, there are no O stars with T$_{eff}$ $\sim$ 40kK at 
luminosities lower than log $L/L_{\odot}$ < 5.4. Also, T$_{eff}$'s about 
20kK are reached by B supergiants, roughly at B1-2I spectral types, which are 
not present in our sample. 

Hence, we do not have several stars in the parameters range where 
the $D_{mom}$ dependence on T$_{eff}$ was observed by \citet{VinkSander21}.
New calculations would be needed to address this issue. 
Alternatively, a careful comparison of a large sample of stars with 
specific spectral types would be desirable. For example, B1-2I and
O5V stars have T$_{eff}$ about 20kK and 40kK, respectively, and similar
luminosities (log $L/L_{\odot} \sim 5.5$). On average, their modified wind-momenta
should be different by $\sim$0.5 dex, according to the theoretical results.

In conclusion, recent theoretical calculations obtain what is expected for line 
driven winds, but still do not agree on the exact WLR and mass-loss 
rate dependence on metallicity. Our empirical WLRs are not matched, a fact 
which is probably linked to the weak wind problem and the very fast winds 
reported. Interestingly, these very fast winds seem to be a common feature 
of these recent theoretical solutions and should be carefully tested against 
multi-wavelength observations. 

\subsubsection{CAK-theory and the WLRs}

It is also interesting to compare our results, in particular the slopes of the WLRs 
obtained, with what is expected from the CAK equations. 
From first principles, the total radiative line acceleration is a 
sum of the contribution by optically thin and optically thick lines 
($g_{rad}^{lines} = \sum_{i} g_i^{thin} + \sum_{i} g_i^{thick}$). 
The respective expressions are very different from one another, with 
the acceleration due to optically thick lines depending on the gradient of 
the velocity field.

In order to solve the hydrodynamics, these accelerations must be known 
at each wind depth point. This can be done by the use of a line-strength
 distribution function -- in terms of an $\alpha$ power-law -- usually 
 calculated from the opacities of thousands to millions of line transitions. 
 The resulting expression is:
 
 \begin{equation}
     g_{rad}^{lines} =  \sum_i g_i^{thin} + \sum_i g_i^{thick} \propto N_{eff}L \left( \frac{dv/dr}{\rho} \right)^\alpha
 \end{equation}

where $N_{eff}$ is the effective number of lines that drives the wind, $\rho$ the gas density and $dv/dr$ the 
velocity gradient \citep[see][]{Kudritzki00,Puls00}. With this acceleration (neglecting rotation), the hydrodynamics 
solution provides expressions for the velocity field and mass-loss rate, as a function of 
$\alpha'=\alpha + \delta$, where $\delta$ is the ionization parameter. The resulting mass-loss rate is $\dot{M} \propto N_{eff}^{1/\alpha'}L^{1/\alpha'}(M(1-\Gamma))^{1-1/\alpha'}$, where $\Gamma$ is the Eddington factor 
and $V_\infty \sim 2.25\frac{\alpha}{1-\alpha}V_{esc}$. The mass-loss depends not only on the number 
of effective lines but also on the luminosity and mass. However, 
the expression for the logarithm of the modified momentum ($\dot{M}V_\infty\sqrt{R/R_\odot}$), when $\alpha'$ is 
relatively close to $2/3$, returns a weak or no mass dependence. Hence:

\begin{equation}
    \label{cakdmom}
    \log D_{mom} \sim \frac{1}{\alpha'} \log L + \text{const.}
\end{equation}

Therefore, our inferred slopes -- $\beta$'s --  can be directly compared with $1/\alpha'$.
From Table \ref{tab:linearcoeffs}, we can infer that\footnote{Do not confuse the $\alpha$'s used in this section with the linear coefficients listed in Table \ref{tab:linearcoeffs}}:

\begin{equation}
  \left( \frac{1}{\alpha'}  \right)_{SMC} <  \left( \frac{1}{\alpha'}  \right)_{MW}   
\end{equation}
  
that is, $\alpha_{MW} < \alpha_{SMC}$, neglecting $\delta$ ($\sim 0.1$). This leads to the conclusion 
that the radiative acceleration of the wind of SMC stars is higher than in the MW stars. 
By recalling the metallicities in these two environments, the contrary is expected.
Thus, our results are in contradiction with the CAK-theory. However, this is not so surprising. 

We note that the empirical data points at low luminosities  
have of course a huge influence on the derived slopes (see Fig. \ref{fig:MWSMC}). These same points confront the radiatively driven wind theory, 
as we already discussed (weak wind problem; see Sect. \ref{sec:discussion}). 
There is the possibility that these winds are not entirely CAK-driven, as 
been already pointed out in the literature \citep[e.g.,][]{Muijres12}. Other physical mechanisms in 
addition to radiation pressure might be at play, invalidating the direct use of the equations above for simple estimates.

As a matter of fact, if we neglect the low luminosity part of our WLR and focus on stars with $\log L/L_\odot \gtrsim 5.4$, 
we find a lower slope for the MW ($\beta_{MW} \sim 1.7$) and the SMC ($\beta_{SMC} \sim 2.2$), 
satisfying $\alpha_{MW} > \alpha_{SMC}$ and thus what is expected in terms of line-statistics. Interestingly, 
in this case not only the slopes but the respective linear coefficients found ($\sim 19.7$ for the MW and $\sim 16.0$ for the SMC) are in reasonable agreement with the ones found by \citet{Mokiem07}, within the error bars (see Table \ref{tab:linearcoeffs}).

The low L end of the WLR brings another difficulty. The detailed analysis on line-statistics presented by \citet{Puls00} indicates that $\alpha$ decreases with T$_{eff}$, in thin winds and/or low Z environments. These conditions are met exactly at low L in our sample. Indeed, our empirical slopes imply $\alpha \sim 1/4$. 
This was also previously reported by \citet{Martins05}, from their sample of O dwarfs. From a theoretical point of view however, when $\alpha$ values are very distinct from 2/3, Eqn. \ref{cakdmom} can present an important mass term, which is neglected. 

Overall, it can be said that our WLRs follow well the CAK theory equations for objects with dense winds, i.e., the brighter objects. However, difficulties and contradictions arise when low L objects are considered. 


\section{Conclusions}
\label{sec:conclusions}

We gathered empirical data of several Milky Way and Small Magellanic O and B stars 
of several spectral classes. Most of them were analyzed by our group along the years, 
using the CMFGEN code. We analyzed the data to address the empirical metallicity 
dependence of wind properties. The main results of our paper are summarized 
below:

\begin{itemize}

    \item Based on CMFGEN models that consistently fit the UV and optical spectra of massive OB stars, 
    we found a clear dependence of the wind strengths on luminosity and metallicity, as expected by the radiatively 
    driven wind theory.

    \item We improved the analysis presented by \citet{Mokiem07}, regarding MW x SMC stars. 
    We analyzed a large luminosity range, in contrast with their work, and explored the 
    luminosity dependence of the Z-dependence of the mass-loss rate. We used our results to 
    estimate and visualize the $m$ values of the $\dot{M} \sim $Z$^m$ relation, which is widely 
    used in the literature. We found $m$ values of $\sim 0.5 - 0.8$, when $\log L/L_\odot$ 
    is $\gtrsim$ 5.4 (i.e., bright objects). However, the Z-dependence seems to get weaker 
    at low $L$. If confirmed by more data, this last finding has likely important astrophysical 
    consequences (e.g., angular momentum evolution, mixing).

    \item We also analyzed wind strengths giving weight to H$\alpha$ mass-loss rate measurements.
    The Z dependence remains, but it does not get weaker at low $L$. We discussed  
    that H$\alpha$ is not free from uncertainties (e.g., contamination, line filling) and there is no 
    reason to favor H$\alpha$ over UV measurements. It is more reliable to consider results that are 
    based on consistent fits to these two spectral regions.

    \item The terminal velocities of the stars of our sample also suggest a Z-dependence. 
    SMC stars seem to have lower terminal wind speeds than MW stars, on average.
    We estimated the $n$ exponent in the $V_{\infty} \sim $Z$^n$ relation to be $\sim 0.1 - 0.2$, 
    further supporting the widely used relation provided long ago by \citet{Leitherer92}. However, 
    a larger sample should be analyzed to confirm our results, as discussed in Section \ref{sec:vinfz}.

    \item Our derived wind strengths for SMC stars are in good agreement with the ones inferred recently 
    by \citet{Ramachandran19}, based on an independent radiative transfer code (PoWR). 
    This brings support to the WLR obtained with CMFGEN and the inferred Z-dependence.

    \item Mass-loss rate and $\log D$ measurements based on bow shocks around O and B stars in the MW \citep{Kobul19}  
    present a large scatter at low luminosities. In the WLR, they fall mostly above the 
    CMFGEN results. The large range in $\dot{M}$ values for B0-O9V stars obtained by this method should be 
    further investigated, since they are likely in contrast to the fairly uniform UV wind features of these objects.
    
    \item Theoretical calculations carried out by different authors find the metallicity dependence of the mass-loss rates of massive stars, but 
    the values reported are not precise. The $m$ exponents (in $\dot{M}$(Z) $\sim$ Z$^m$) vary considerably, but are close to our results. Part of the discrepancy between our results and theory at low L is likely linked to 
    the weak wind problem and the very high terminal velocities predicted (e.g., greatly exceeding 3000 km s$^{-1}$).
    Despite the complexities and known limitations of NLTE expanding atmosphere models, the data provided here have an empirical nature and should serve as a guide for future comparisons.

    \item We compared the slopes of our WLRs to what is expected in terms of the CAK equations. Regarding line-statistics, contradictory results are obtained due the low luminosity stars of our sample -- $\alpha_{SMC} > \alpha_{MW}$. When this region is neglected, no contradictions arise and our WLRs are in good agreement with the ones previously obtained by \citet{Mokiem07}. We attribute the issue at low L to the weak wind problem and also a possible misuse of the CAK equations. Other physical mechanisms in addition to radiation pressure might be at play, invalidating the direct use of CAK equations for simple estimates.

\end{itemize}

It would be helpful to increase the sample of analyzed stars using optical 
and UV data, consistently. In this context, the Hubble UV Legacy 
Library of Young Stars as Essential Standards (ULLYSES) will be an excellent opportunity 
in the next years. Robust mass-loss rates for Magellanic Clouds stars would 
provide better constrains on the relations provided in this work. 

\section*{Acknowledgements}

WLFM acknowledges CNPq for the PQ grant (307152/2016-2) and John Hillier for making CMFGEN available to the massive star community.
MBP gratefully acknowledges funding from the German \emph{Deut\-sche For\-schungs\-ge\-mein\-schaft, DFG\/} in the form of an Emmy Noether Research Group (grant number SA4064/1-1, PI Sander). 

\section*{DATA AVAILABILITY}
The data underlying this article are available in the article and in its online supplementary material.










\bsp	
\label{lastpage}
\end{document}